\documentclass[journal, 10pt]{IEEEtran}

\usepackage{multirow}
\usepackage[font=scriptsize,caption=false,labelsep=space]{subfig}
\usepackage{mathrsfs}
\usepackage{graphicx,float,wrapfig,epstopdf,amsmath}
\usepackage[]{algorithmicx}
\usepackage{algpseudocode,algorithm}
\usepackage{balance}
\usepackage{mathtools}
\usepackage{amsmath}
\usepackage{amssymb}
\usepackage{amsthm}
\usepackage{graphicx}
\usepackage{epstopdf}
\usepackage{times}
\usepackage{textcomp,cite}
\usepackage{xcolor}
\usepackage{url}
\usepackage{hyperref}

\usepackage{multirow}
\usepackage{threeparttable}
\graphicspath{{./figures/}}

\DeclareMathOperator*{\minimize}{minimize} 
\DeclareMathOperator*{\subjectto}{subject \hspace{3pt} to:\hspace{3pt}} 

\begin{document}
	
	\title{ Federated Learning for Channel Estimation in Conventional and RIS-Assisted Massive MIMO
	}
	\author{Ahmet~M.~Elbir\textit{, Senior Member, IEEE}, and Sinem Coleri\textit{, Senior Member, IEEE} 
		
		\thanks{	The work of Sinem Coleri was supported by the Scientific and Technological Research Council of Turkey with European CHIST-ERA grant 119E350.}
		
		\thanks{A. M. Elbir is with the Department of Electrical and Electronics Engineering, Duzce University, Duzce, Turkey, and with University of Hertfordshire, Hatfield, UK  (e-mail: ahmetmelbir@gmail.com).} 
		\thanks{S. Coleri is with the Department of Electrical and Electronics Engineering, 
			Koc University, Istanbul, Turkey (e-mail: scoleri@ku.edu.tr).}
	}
	\maketitle
	
	\begin{abstract}
		Machine learning (ML) has attracted a great research interest for physical layer design problems, such as channel estimation, thanks to its low complexity and robustness. Channel estimation via ML requires model training on a dataset, which usually includes the received pilot signals as input and channel data as output. In previous works, model training is mostly done via centralized learning (CL), where the whole training dataset is collected from the users at the base station (BS). This approach introduces huge communication overhead for data collection. In this paper, to address this challenge, we propose a federated learning (FL) framework for channel estimation. We design a convolutional neural network (CNN) trained on the local datasets of the users without sending them to the BS. We develop FL-based channel estimation schemes for both conventional and RIS (intelligent reflecting surface) assisted massive MIMO (multiple-input multiple-output) systems, where a single CNN is trained for two different datasets for both scenarios. We evaluate the performance for noisy and quantized model transmission and show that the proposed approach provides approximately 16 times lower overhead than CL, while maintaining satisfactory performance close to CL. Furthermore, the proposed architecture exhibits lower estimation error than the state-of-the-art ML-based schemes.
	\end{abstract}
	\begin{IEEEkeywords}
		Channel estimation, Federated learning, Machine learning, Centralized learning, Massive MIMO.
	\end{IEEEkeywords}

	\section{Introduction}
	\label{sec:Introduciton}
	Compared to the cellular communication systems in lower frequency bands, millimeter wave (mm-Wave) signals, with the frequency range $30$-$300$ GHz, encounter a more complex propagation environment that is characterized by higher scattering, severe penetration losses, and higher path loss for fixed transmitter and receiver gains~\cite{mimoOverview,mimoScalingUp,5GwhatWillItBe}. These losses are compensated by providing beamforming power gain through massive number of antennas at both transmitter and receiver with multiple-input-multiple-output (MIMO) architecture. However, such a large antenna array requires a dedicated radio-frequency (RF) chain for each antenna, resulting in an expensive system architecture and high power consumption. In order to address this issue and reduce the number of digital RF components, hybrid analog and baseband beamforming architectures have been introduced, wherein a small number of phase-only analog beamformers are employed~\cite{mimoRHeath}.
	As a result, the combination of high-dimensional analog and low-dimensional baseband  beamformers significantly reduces the number of RF chains while  maintaining sufficient beamforming gain \cite{mimoRHeath}.
	
	Even with the reduced number of RF chains, the hybrid beamforming architecture combined with mm-Wave transmission comes with the expensive cost of energy consumption and hardware complexity~\cite{lis_COMmag}.  In order to address these issues and provide a more green and suitable solution to enhance the wireless network performance, reconfigurable intelligent surfaces (RISs) (also known as   intelligent reflecting surfaces) are envisaged as a promising solution with low cost and complexity~\cite{elbir2020DL4RIS_survey,irs_holographic,lis_COMmag,lis_2018_TWC,irsChongwen}. An RIS is an electromagnetic 2-D surface that is composed of large number of passive reconfigurable meta-material elements, which reflect the incoming signal by introducing a pre-determined phase shift. This phase shift can be controlled via external signals by the base station (BS) through a backhaul control link. As a result, the incoming signal from the BS can be manipulated in real-time, thereby, reflecting the received signal towards the users. Hence, the usage of RIS improves the received signal energy at the distant users as well as expanding the coverage of the BS.

	In both conventional and RIS-assisted massive MIMO scenarios, the performance of the system architecture strongly relies on the accuracy of the instantaneous channel state information (CSI), given the highly dynamic nature of the mm-Wave channel \cite{coherenceTimeRef}. Thus, the channel estimation accuracy plays an important role in the design of  the analog and digital beamformers in  conventional massive MIMO~\cite{alkhateeb2016frequencySelective,sohrabiOFDM}, and the design of reflecting beamformer phase shifts of the  RIS elements in the RIS-assisted scenario~\cite{lis_channelEst_reflectedBFDesign,lis_2018_TWC}. Furthermore,  RIS-assisted massive MIMO involves signal reception through multiple channels (e.g., BS-RIS, RIS-user and BS-user), which makes the channel estimation task more challenging and interesting. As a result, several channel estimation schemes are proposed for massive MIMO and RIS-assisted scenarios, based on 	compressed sensing~\cite{lis_channelEst_reflectedBFDesign}, angle-domain processing~\cite{mimoAngleDomainFaiFai} and coordinated pilot-assignment~\cite{channelEstLargeArrays2}. The performance of these analytical approaches strongly depends on the perfection of the antenna array output so that reliable channel estimation accuracy can be obtained. In order to provide robustness against the imperfections/corruptions in the array data, data-driven techniques, such as machine learning (ML) based approaches, have been proposed to uncover the non-linear relationships in data/signals with lower computational complexity and achieve better performance for parameter inference and be tolerant against the imperfections in the data. {\color{black}As listed below, ML is more efficient than model-based techniques that largely rely on mathematical models:
		\begin{itemize}
			\item A learning model constructs a non-linear mapping between the raw input data and the desired output to approximate a problem from a model-free perspective. Thus, its prediction performance is robust against the corruptions/imperfections in the wireless channel data.
			\item ML learns the feature patterns, which are easily updated for the new data and adapted to environmental changes. In the long run, this results in a lower computational complexity than the model-based optimization.
			\item ML-based solutions have significantly reduced run-times because of parallel processing capabilities. On the other hand, it is not straightforward to achieve parallel implementations of conventional optimization and signal processing algorithms. 
		\end{itemize}
	}

	\begin{figure}[t]
		\centering
		\subfloat[]{\includegraphics[width=.8\columnwidth]{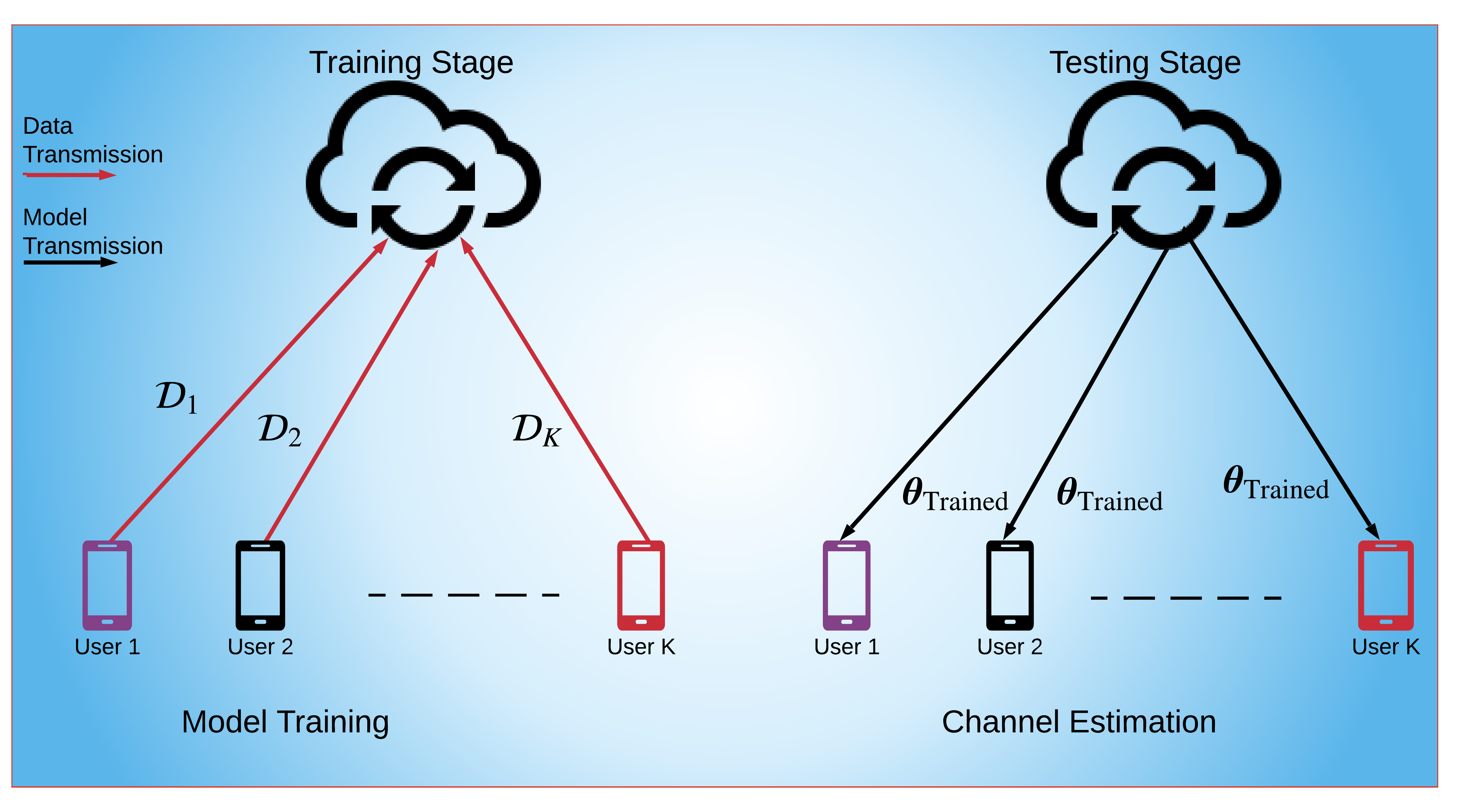} \label{fig_DiagramML}}\\
		\subfloat[]{\includegraphics[width=.8\columnwidth]{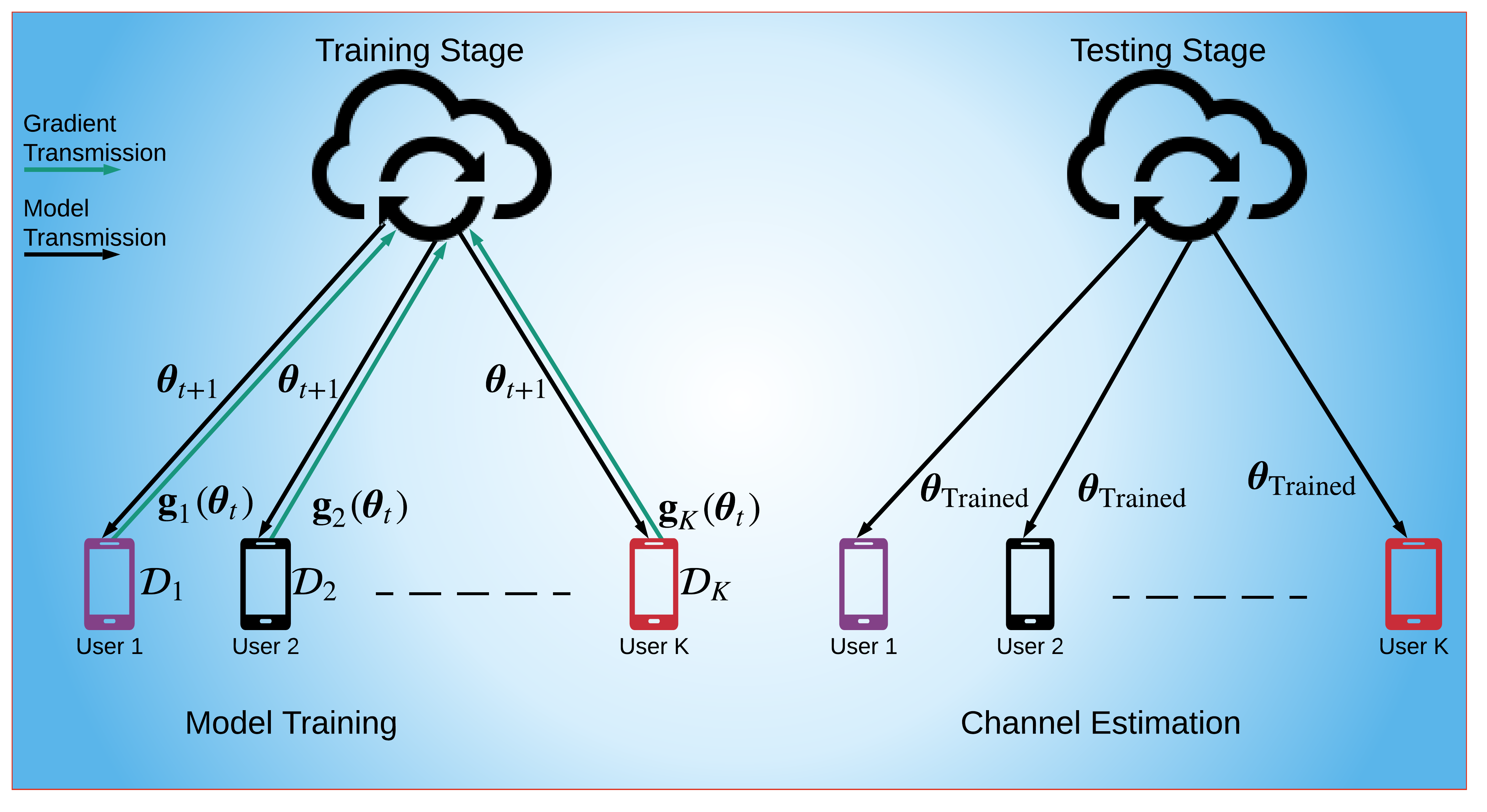}\label{fig_DiagramFL} }
		\caption{Model training and testing stages for (a) CL and (b) FL. \color{black}During training, CL involves the transmission of datasets $\mathcal{D}_{k\in \mathcal{K}}$ from the users to the server, while users send only the model updates $\mathbf{g}_{k\in \mathcal{K}}(\boldsymbol{\theta}_t)$ in FL. In the test stage of both CL and FL, the server broadcasts the trained learning models $\boldsymbol{\theta}_\mathrm{Trained}$ to the users.  }
		\label{fig_Diagram}
	\end{figure}

	In massive MIMO and RIS-assisted systems,  ML has been proven to have higher spectral efficiency and lower computational complexity for the problems such as channel estimation~\cite{deepCNN_ChannelEstimation,deepLearningChannelAndDOAEst,elbir_LIS}, hybrid beamforming~\cite{elbirDL_COMML,elbirQuantizedCNN2019,elbirHybrid_multiuser} and angle-of-arrival (AoA) estimation~\cite{elbirIETRSN2019,elbir_DL_MUSIC}.

	In ML context, channel estimation problem is solved by training a model, e.g., a neural network (NN), on the local datasets collected by the users~\cite{deepCNN_ChannelEstimation,deepLearningChannelAndDOAEst,elbir_LIS}. The trained model provides a non-linear mapping between the input data, which can be usually selected as the received pilot signals, and the output data, i.e., the channel data. Previous works mostly consider centralized learning (CL) schemes where the whole dataset, i.e., input-output data pairs, is transmitted to the BS \textcolor{black}{(via RIS in the RIS-assisted scenario)} for model training, as illustrated in Fig.~\ref{fig_DiagramML}. Once the model is trained at the BS, then the model parameters are sent to the users, which can perform channel estimation task by feeding the model with the received pilot data. However, this approach involves huge  communication overhead, i.e., transmitting the whole dataset from users to the BS. For example, in  LTE (long term evolution), a single frame of $5$ MHz bandwidth and $10$ ms duration can carry only $6000$ complex symbols~\cite{FL_gunduz_fading}, whereas the size of the whole dataset can be on the order hundreds of thousands symbols~\cite{elbirHybrid_multiuser,elbirQuantizedCNN2019,deepCNN_ChannelEstimation,deepLearningChannelAndDOAEst}. As a result, CL-based techniques demand huge bandwidth requirements. 
	
	In order to deal with high communication overhead of CL schemes, recently federated learning (FL) schemes have been proposed~\cite{spm_federatedLearning,elbir2020federated}. In FL, instead of sending the whole dataset, only the model updates, i.e., gradients of the model parameters, are transmitted, as illustrated in Fig.~\ref{fig_DiagramFL}. As a result, the communication overhead is reduced. In the literature, FL has been considered for the scheduling and power allocation in wireless sensor networks~\cite{FL_Bennis3}, the trajectory planning of UAV (unmanned aerial vehicle) networks~\cite{FL_Bennis2}, task fetching and offloading in vehicular networks~\cite{FL_Bennis5,elbir2020federated}, image classification in~\cite{FL_gunduz_fading,FL_Gunduz}, and massive MIMO hybrid beamforming design~\cite{elbir2020FL_HB}. All of these studies accommodate  multiple edge devices exchanging model updates with the parameter server to train a global model.	
	In the aforementioned works, FL has been mostly used for image classification/object detection problems in different networking schemes by the assumption that the perfect CSI is available. Motivated by the fact that the acquisition of CSI is very critical in massive MIMO systems and FL has not been considered directly for the channel estimation problem, in this work, we leverage FL for the channel estimation problem, which has been studied previously in the context of  CL-based training~\cite{deepCNN_ChannelEstimation,deepLearningChannelAndDOAEst,elbir_LIS,elbir2019online}. \textcolor{black}{Compared to CL, FL is more applicable in case of distributed devices, such as mobile phones. Furthermore, training the same model with FL, rather than CL, reduces the communication overhead significantly during training while maintaining satisfactory channel estimation performance close CL.} To the best of our knowledge, this is the first work for the use of FL in channel estimation.

	In this paper, we propose an FL-based model training approach for channel estimation problem in both conventional and RIS-assisted massive MIMO systems. We design a convolutional neural network (CNN), which is located at the BS and trained on the local datasets. \textcolor{black}{For these datasets, where the input is received pilot signal and the output is the channel matrix, the usage \textcolor{black}{of the} CNN is more convenient than the recurrent NNs (RNNs), which are designed to predict the future CSI by using the previous channels based on the sequential data~\cite{rnnChannelEst2_channelAging}.} The proposed approach has three stages, namely, data collection, training and prediction. In the first stage, each user collects its training datasets and stores them for model training, which is not explicitly discussed in the previous ML-based works~\cite{deepCNN_ChannelEstimation,deepLearningChannelAndDOAEst,elbir_LIS,mimoDLChannelEstimation}.  In the second stage, each user uses its own local dataset, and computes the model updates and sends them to the BS\textcolor{black}{\footnote{\color{black}The model parameters computed at the users are transmitted to the BS via the RIS in RIS-assisted scenario.}}, where the model updates are aggregated to train a global model.  \textcolor{black}{The main advantage of the proposed FL approach is the reduction in the communication overhead. This overhead is proportional to the dimensionality of the channel matrix, which can be higher in RIS-assisted systems than the conventional MIMO due to the large number of RIS elements.} Apart from that, the proposed approach reduces the computation time as well as increasing the robustness against data corruptions. \textcolor{black}{One of the main challenges in FL-based channel estimation is due to the non-i.i.d. (independent identical distribution) structure of the training data. FL is known to converge faster if the local datasets are i.i.d.~\cite{fl_By_Google}. Since the channel estimation dataset is non-i.i.d. because of the distribution of the user locations, FL is expected to converge slower. In order to improve the performance in non-i.i.d. scenario, using deeper and wider learning models help to provide better feature extraction and representation performance~\cite{elbir2020FL_HB}. Thus, we perform a hyper-parameter optimization to achieve a satisfactory performance.}

	
	The main contributions of this paper can be summarized as follows:
	\begin{enumerate}
		\item We propose an FL-based channel estimation approach for both conventional and RIS-assisted massive MIMO systems. Different from the conventional centralized model learning techniques, the proposed FL framework provides decentralized learning, which	
		significantly reduces the communication overhead compared to the CL-based techniques while maintaining satisfactory channel estimation performance close to CL. 
		\item  In order to estimate both direct (BS-user) and cascaded (BS-RIS-user) channels in RIS-assisted scenario, input and output data are combined together for each communication link, hence a single CNN architecture is designed, instead of using different NNs for each task.
		
		%
		%

		\item  We \textcolor{black}{prove the convergence of FL} and demonstrate its superior performance over CL in terms of communication overhead and channel estimation accuracy  via extensive numerical simulations for different number of users while considering the quantization and corruption of the gradient and model data as well as the loss of  a portion of the model data during transmission. 
	\end{enumerate}
	
	%

	Throughout the paper, the identity matrix of size $N\times N$ is denoted by $\mathbf{I}_N$. $(\cdot)^T$ and $(\cdot)^H$ denote transpose and  conjugate transpose operations, respectively. For a matrix $\mathbf{A}$ and a vector $\mathbf{a}$, $[\mathbf{A}]_{i,j}$ and $[\mathbf{a}]_i$ denote the $(i,j)$th element of matrix $\mathbf{A}$ and the $i$th element of vector $\mathbf{a}$, respectively.  The function $\mathbb{E}\{\cdot\}$ provides the statistical expectation of its argument and  $\angle \{\cdot\}$ measures the angle of complex quantity. $\|\mathbf{A}\|_\mathcal{F}$ and $\|\mathbf{a}\|_2$ denote the Frobenius and $l_2$-norm, respectively. $\otimes$ is the Hadamard element-wise multiplication and $\nabla_{\mathbf{a}}$ represents the gradient with respect to $\mathbf{a}$. A convolutional layer with $N$ $D\times D$ 2-D kernel is represented by $N$@ $D\times D$.

	\section{System Model}
	\label{sec:SystemModel}
	We consider a multi-user MIMO-OFDM (orthogonal frequency division multiplexing) system with $M$ subcarriers, where the BS has $N_\mathrm{BS}$ antennas to communicate with $K$ users, each of which has $N_\mathrm{MS}$ antennas. In the downlink, the BS first precodes $K$ data symbols $\mathbf{s}[m] = [s_1[m],s_2[m],\dots,s_{K}[m]]^\textsf{T}\in \mathbb{C}^{K}$ at each subcarrier ($m\in \mathcal{M} = \{1,\dots,M\}$) by applying the subcarrier-dependent baseband precoders $\mathbf{F}_{\mathrm{BB}}[m] = [\mathbf{f}_{\mathrm{BB}_1}[m],\mathbf{f}_{\mathrm{BB}_2}[m],\dots,\mathbf{f}_{\mathrm{BB}_{K}} [m]]\in \mathbb{C}^{K\times K}$. Then, the signal is transformed to the time-domain via $M$-point inverse discrete Fourier transform (IDFT). After adding cyclic prefix (CP), the BS employs subcarrier-independent analog precoder $\mathbf{F}_\mathrm{RF}\in \mathbb{C}^{N_\mathrm{BS}\times K}$ to form the transmitted signal.  Given that $\mathbf{F}_{\mathrm{RF}}$ consists of analog phase shifters, we assume that the RF precoder has constant unit-modulus constraints, i.e., $|[\mathbf{F}_{\mathrm{RF}}]_{i,j}|^2 =1$. Additionally, we have the power constraint  $\sum_{m=1}^{M}\|\mathbf{F}_{\mathrm{RF}}\mathbf{F}_{\mathrm{BB}}[m] \|_\mathcal{F}^2= MK$ that is enforced by the normalization of the baseband precoder $\{\mathbf{F}_{\mathrm{BB}}[m] \}_{m\in \mathcal{M}}$. Thus, the transmitted signal becomes $	\mathbf{x}[m] = \mathbf{F}_\mathrm{RF}\sum_{k=1}^{K} \mathbf{f}_{\mathrm{BB}_k}[m] s_k[m].$
	\subsection{Channel Model}
	Before reception at the users, the transmitted signal is passed through the mm-Wave channel, which can be  represented by a geometric model with limited scattering \cite{alkhateeb2016frequencySelective}. Let us define $\mathbf{H}_k[m]$ as the $N_\mathrm{MS}\times N_\mathrm{BS}$ mm-Wave channel matrix between the BS and the $k$th user. Then, $\mathbf{H}_k[m]$ includes the contributions of $L$ paths, each of which has the time delay $\tau_{k,l}$ with  relative  AoA $\bar{\phi}_{k,l} \in \Theta$ ($\Theta = [-\frac{\pi}{2},\frac{\pi}{2}]$), angle-of-departure (AoD) $\phi_{k,l} \in\Theta$, and the complex path gain $\alpha_{k,l}$ for the $k$th user and $l$th path. Let $p(\tau)$ denote a pulse shaping function for $T_\mathrm{s}$-spaced signaling evaluated  at $\tau$ seconds. Then, the mm-Wave delay-$d$ MIMO channel matrix in time domain is given by
	\begin{align}
	\label{eq:delaydChannelModel}
	\bar{\mathbf{H}}_k[d] = &  \sqrt{\frac{ N_\mathrm{BS} N_{\mathrm{MS}} } {L}} \sum_{l=1}^{L} \alpha_{k,l} p(dT_\mathrm{s} - \tau_{k,l})
	\mathbf{a}_\mathrm{MS}(\bar{\phi}_{k,l}) \mathbf{a}_\mathrm{BS}^\textsf{H}(\phi_{k,l}),
	\end{align}
	where  $\mathbf{a}_\mathrm{MS}(\bar{\phi_{k,l}})$ and $\mathbf{a}_\mathrm{BS}(\phi_{k,l})$ are the $N_\mathrm{MS} \times 1$, and $N_\mathrm{BS}\times 1$ steering vectors representing the array responses of the antenna arrays at the users and the BS, respectively. 
	Let $\lambda_m = \frac{c_0}{f_m}$ be the wavelength for the subcarrier $m$ at frequency  $f_m$. Since the operating frequency is relatively higher than the bandwidth in mm-Wave systems and the subcarrier frequencies are close to each other (i.e., $f_{m_1} \approx f_{m_2}$, $m_1,m_2 \in\mathcal{M}$), we use a single operating wavelength $\lambda = \lambda_{1} = \dots = \lambda_{M} = \frac{c_0}{f_c}$, where $c_0$ is speed of light and $f_c$ is the central carrier frequency \cite{alkhateeb2016frequencySelective,sohrabiOFDM}. This approximation also allows for a single frequency-independent analog beamformer for each subcarrier.  Then, for a uniform linear array (ULA), the array response of the antenna array at the BS is
	\begin{align}
	\mathbf{a}_\mathrm{BS}(\phi) = \big[ 1, e^{j\frac{2\pi}{\lambda} {d}_\mathrm{BS}\sin(\phi)},\dots,e^{j\frac{2\pi}{\lambda} (N_\mathrm{BS}-1){d}_\mathrm{BS}\sin(\phi)} \big]^\textsf{T},
	\end{align}
	where ${d}_\mathrm{BS} = \lambda/2$ is the antenna spacing. The $n$th element of $\mathbf{a}_\mathrm{MS}(\bar{\phi})$ can be defined in a similar way as for $\mathbf{a}_\mathrm{BS}(\phi)$ as $\left[\mathbf{a}_\mathrm{MS}(\bar{\phi})\right]_n = e^{j\pi (n-1)\sin(\bar{\phi})}$, $n = 1,\dots, N_\mathrm{MS}$. 
	{\color{black} After performing $M$-point DFT of the delay-$d$ channel model in (\ref{eq:delaydChannelModel}), the channel matrix  of the $k$th user at subcarrier $m$ becomes
		\begin{align}
		\label{Hm_OFDM}
		\mathbf{H}_k[m] = \sum_{d=0}^{D-1}\bar{\mathbf{H}}_k[d]e^{-j\frac{2\pi m}{M} d},
		\end{align}
		where $D\leq M$ is the CP length. The frequency domain channel in (\ref{Hm_OFDM}) is used in MIMO-OFDM systems, where the orthogonality of each subcarrier is held such that  $||\mathbf{H}_k^\textsf{H}[m_1]\mathbf{H}_k[m_2]||_\mathcal{F}^2 = 0 $ for $m_1, m_2 \in \mathcal{M}$ and $m_1 \neq m_2$.}

	With the aforementioned block-fading channel model \cite{alkhateeb2016frequencySelective}, the received signal at the $k$th user before analog processing at subcarrier $m$ is $	\tilde{\mathbf{y}}_k[m] =\sqrt{\rho} \mathbf{H}_k[m] \mathbf{x}[m] $, i.e.,
	\begin{align}
	\label{arrayOutput}
	\tilde{\mathbf{y}}_k[m] = \sqrt{\rho}\mathbf{H}_k[m] \mathbf{F}_\mathrm{RF}\mathbf{F}_\mathrm{BB}[m]\mathbf{s}[m] + \mathbf{n}[m],
	\end{align}
	where $\rho$ represents the average received power and $\mathbf{n}[m] \sim \mathcal{CN}({0},\sigma^2 \mathbf{I}_\mathrm{N_\mathrm{MS}})$ is additive white Gaussian noise (AWGN) vector. At the $k$th user, the received signal is first processed by the analog combiner $\mathbf{w}_{\mathrm{RF},k}\in \mathbb{C}^{N_\mathrm{MS}}$. Then, the cyclic prefix is removed from the processed signal and $M$-point DFTs are applied to yield the signal in frequency domain.
	Then, the received baseband signal becomes
	\begin{align}
	\label{eq:ReceivedMIMO}
	\bar{y}_k[m] =\sqrt{\rho}\mathbf{w}_{\mathrm{RF},k}^\textsf{H} \mathbf{H}_k[m] \mathbf{F}_\mathrm{RF}\mathbf{F}_\mathrm{BB}[m]\mathbf{s}[m] + \mathbf{w}_{\mathrm{RF},k}^\textsf{H}\mathbf{n}[m],
	\end{align}
	where the analog combiner $\mathbf{w}_{\mathrm{RF},k}$ has the constraint $\big[\mathbf{w}_{\mathrm{RF},k}\mathbf{w}_{\mathrm{RF},k}^\textsf{H}\big]_{i,i}=1$, similar to the RF precoder. Once the received symbols, i.e., $	y_k[m]$ are obtained at the $k$th user, they are demodulated according to its respective modulation scheme, and the information bits are recovered for each subcarrier. \textcolor{black}{To accurately recover the data streams $\mathbf{s}[m]$ in (\ref{eq:ReceivedMIMO}), the channel matrix $\mathbf{H}_k[m]$ should be estimated. This is usually done by using pilot signals in the preamble stage~\cite{elbir2020withoutCSI,deepCNN_ChannelEstimation}, wherein the  beamformers $\mathbf{F}_\mathrm{RF}$,  $\mathbf{F}_\mathrm{BB}$ and $\mathbf{w}_\mathrm{RF_k}$ are designed accordingly (See Section~\ref{sec:FLmMIMO}). }
	
	
	\subsection{Problem Description}
	The aim in this work is to estimate the channel matrix $\mathbf{H}_k[m]$ via FL, as illustrated in Fig.~\ref{fig_DiagramFL}. To this end, the global NN for channel estimation (henceforth called \textsf{ChannelNet}) located at the BS is trained on the local datasets of the users. Let $\mathcal{D}_k$ denote the local dataset at the $k$th user, containing the input-output pairs $\mathcal{D}_k^{(i)} = (\mathcal{X}_k^{(i)}, \mathcal{Y}_k^{(i)}) $\footnote{The sizes of $\mathcal{X}_k^{(i)}$ and $\mathcal{Y}_k^{(i)}$ depend on the size of the channel matrix, and they are explicitly given in Sec.~\ref{sec:FLmMIMO} and Sec.~\ref{sec:FL_RIS} for conventional and RIS-assisted massive MIMO scenario, respectively.}, $i = 1,\dots, \textsf{D}_k$ and $\textsf{D}_k = |\mathcal{D}_k|$ is the size of the local dataset $\mathcal{D}_k$. Here, $\mathcal{X}_k^{(i)}$ represents the $i$th input data, i.e., the received pilot signals, $\mathcal{Y}_k^{(i)}$ denotes the $i$th output/label data, i.e., the channel matrix, for $k \in \mathcal{K}$, $\mathcal{K} = \{1,\dots,K\}$.	
	Thus, for an input-output pair $(\mathcal{X},\mathcal{Y})$, \textsf{ChannelNet} constructs a non-linear relationship  between the input and the output data as $f(\mathcal{X}|\boldsymbol{\theta}) = \mathcal{Y}$, where $\boldsymbol{\theta}\in \mathbb{R}^P$ denotes the learnable parameters.
	

	\begin{figure}[t]
		\centering
		\subfloat[]{\includegraphics[width=.7\columnwidth]{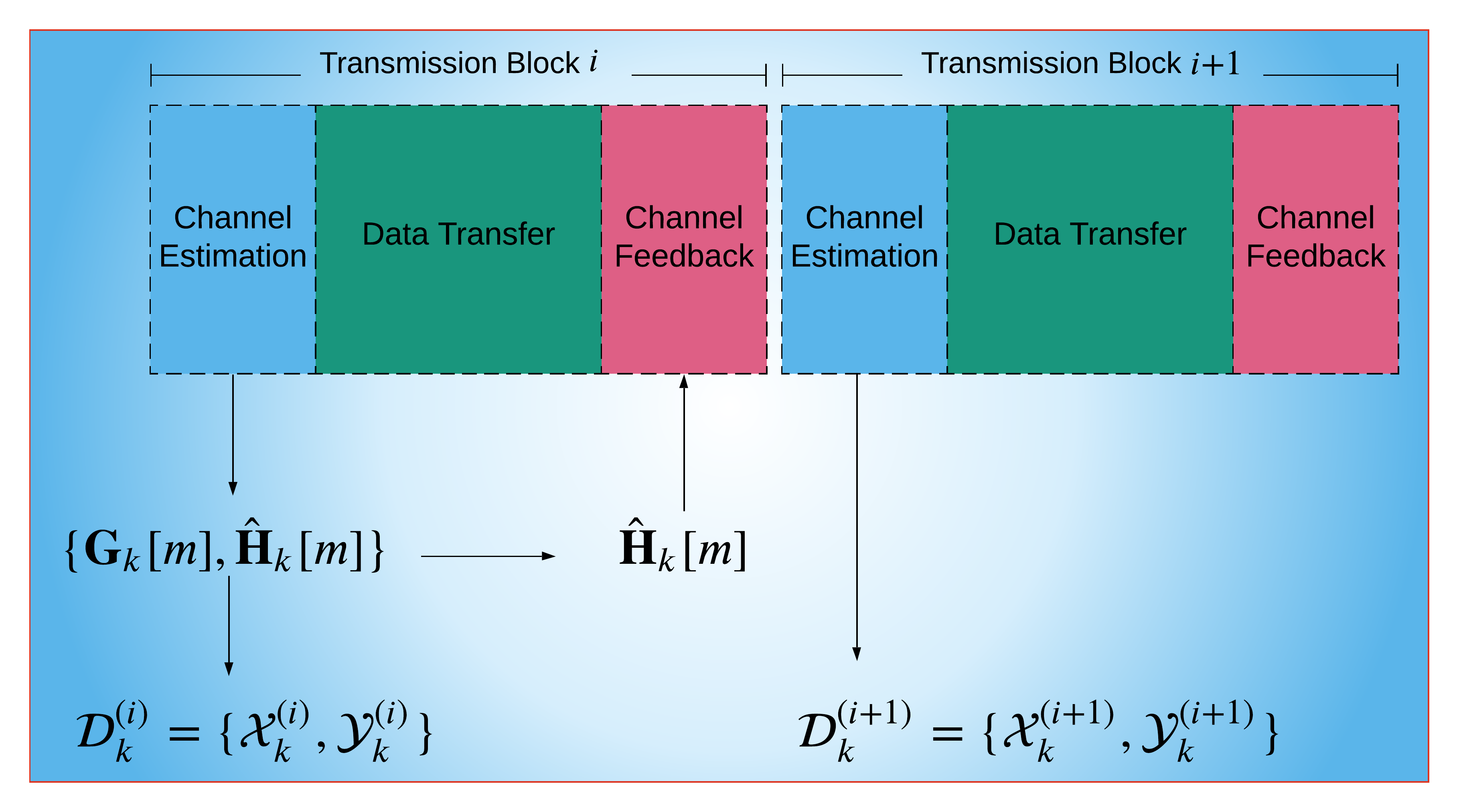} \label{diagDataCollection}}\\
		\subfloat[]{\includegraphics[width=.7\columnwidth]{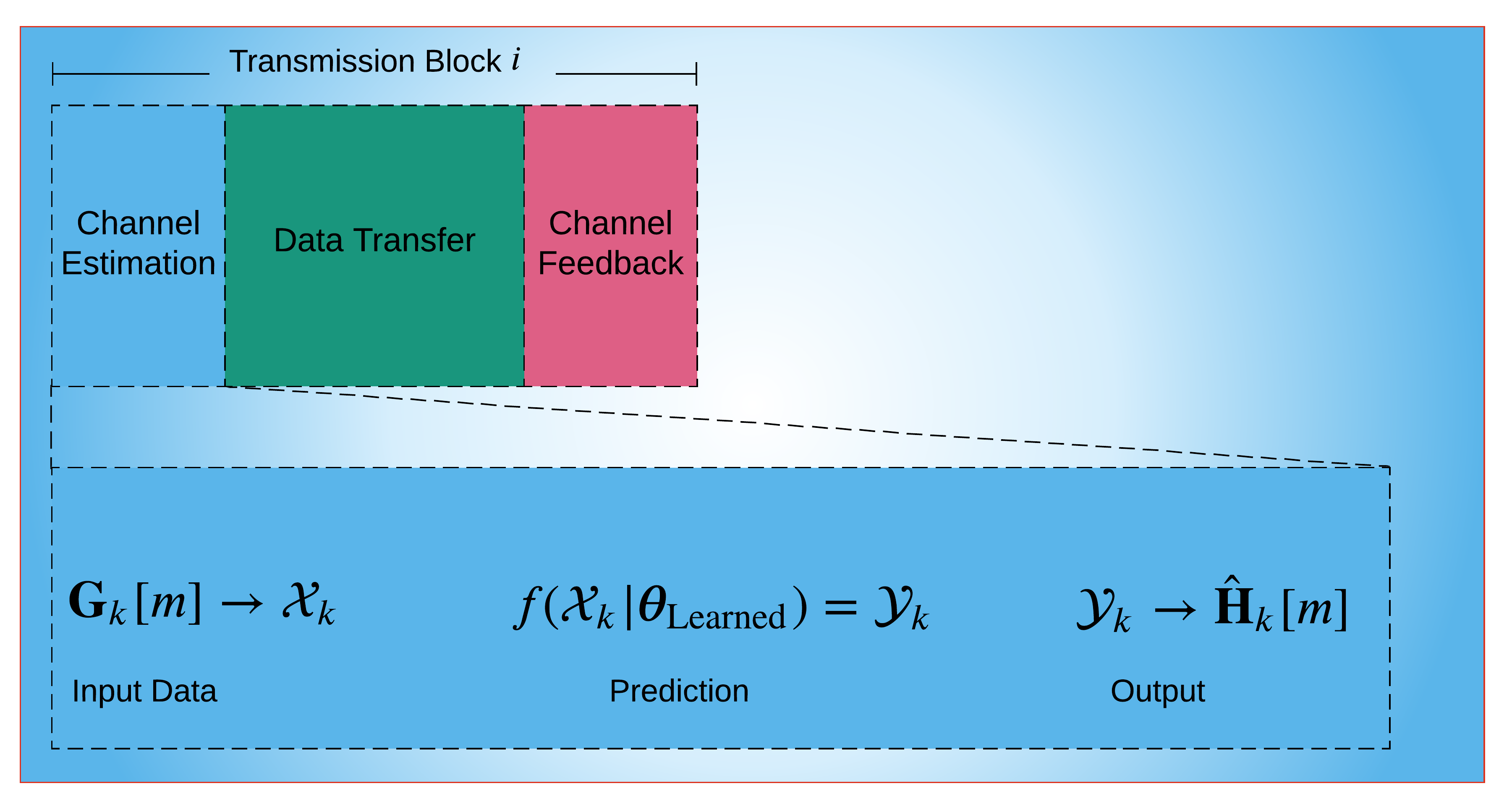} \label{diagPrediction}}
		\caption{(a) Training data collection and (b) channel estimation  with the trained model. }
		\label{fig_DiagDataCollection}
	\end{figure}

	\section{Federated Learning for Channel Estimation}
	\label{sec:FL4CE}
	In this section, we present the proposed FL-based channel estimation scheme, which is comprised of three stages: training data collection, model training and prediction. First, we present the training data collection stage, in which each user collects its own training dataset from the received pilot signals. After providing the FL-based model training scheme, we discuss how the input and output label data are determined for both massive MIMO and RIS-assisted scenarios, respectively. Once the learning model is trained, then it can be used for channel estimation in the prediction stage.
	
	\subsection{Training Data Collection}
	\label{sec:TrainDataCollect}
	In Fig.~\ref{fig_DiagDataCollection}, we present the communication interval at the user for two consecutive data transmission blocks. At the beginning of each transmission block, the received pilot signals are acquired and processed for channel estimation. This can be done by employing one of the analytical channel estimation techniques, which can be based on compressed sensing~\cite{channelEstimation1CS,channelEstLargeArrays}, angle-domain processing~\cite{mimoAngleDomainFaiFai} and coordinated pilot-assignment~\cite{channelEstLargeArrays2}.   The analytical approach is only used in the training data collection stage, which is relatively smaller than the prediction stage~\cite{elbir2019online}. Hence, the use of ML/FL in the prediction stage becomes more advantageous over the analytical techniques in the long term.
	
	It is also worth to mention that the training data can be obtained via offline datasets which are prepared by collecting the data from the field measurements. In~\cite{datasetChannelEst}, authors present a channel estimation dataset, which is obtained by electromagnetic simulations tools. While this approach can also be followed, the offline collected data may not always reflect the channel characteristics and the imperfections in the mm-Wave channel. In this work, we evaluate the performance of the proposed approach on the datasets whose labels are selected as both true and estimated channel data. For the estimated channel, we assume that the training data are collected, as described in Fig.~\ref{fig_DiagDataCollection}, by employing angle-domain channel estimation (ADCE) technique~\cite{mimoAngleDomainFaiFai}, which has close to minimum mean-square-error (MMSE) performance.
	
	After channel estimation, the training data can be collected by storing the received pilot data $\mathbf{G}_k[m]$ and the estimated channel data $\hat{\mathbf{H}}_k[m]$ in the internal memory of the user. (We discuss how $\mathbf{G}_k[m]$ is determined in Sec.~\ref{sec:FLmMIMO}.) Then, the user feedbacks the estimated channel data to the BS via uplink transmission. As a result, the local dataset $\mathcal{D}_k$ can be collected at the $k$th user after  $i = 1,\dots,\textsf{D}_k$ transmission blocks.
	This approach allows us to collect training data for different channel coherence times, which can be very short due to dynamic nature of the mm-Wave channel, such as indoor and vehicular communications~\cite{coherenceTimeRef}.
	
	The above process is the first stage of the proposed FL-based channel estimation framework. Once the training data is collected, the global model is trained (see, e.g., Fig.~\ref{fig_DiagramFL}). After training, each user can estimate its own channel via the trained NN by simply feeding the NN with $\mathbf{G}_k[m]$  and obtains $\hat{\mathbf{H}}_k[m]$, as illustrated in Fig.~\ref{diagPrediction}.


	\subsection{FL-based Model Training}
	\label{sec:FLTraining}
	We begin by introducing the training concept in conventional CL, then develop FL-based model training. 
	
	In CL-based model training for channel estimation~\cite{deepCNN_ChannelEstimation,deepLearningChannelAndDOAEst,elbir2019online,elbir_LIS,mimoDLChannelEstimation}, the training of the global NN is performed by collecting the local datasets $\{\mathcal{D}_k\}_{k\in \mathcal{K}}$ from the users, as illustrated in Fig.~\ref{fig_DiagramML}. Once the BS has collected the whole dataset $\mathcal{D}$, the training is performed by solving the following problem {\color{black}
		\begin{align}
		\label{lossML}
		\minimize_{\boldsymbol{\theta}}  & \hspace{10pt} 
		\mathcal{L}(\boldsymbol{\theta}) \nonumber\\
		\subjectto & \hspace{10pt}f(\mathcal{X}^{(i)}|\boldsymbol{\theta}) = \mathcal{Y}^{(i)}, i = 1,\dots, \textsf{D},
		\end{align}
		where  $\textsf{D} = |\mathcal{D}|$ is the number of training samples and $\mathcal{L}(\boldsymbol{\theta})$ denotes the loss function defined  as
		\begin{align}
		\mathcal{L}(\boldsymbol{\theta}) =  \frac{1}{\textsf{D}}\sum_{i=1}^\textsf{D}\| f( \mathcal{X}^{(i)}|\boldsymbol{\theta}) - \mathcal{Y}^{(i)}  \|_\mathcal{F}^2,
		\end{align}
		which is the MSE between the label data $\mathcal{Y}^{(i)}$ and the prediction of the NN, $f( \mathcal{X}^{(i)}|\boldsymbol{\theta})$.

		On the other hand, in FL, the local datasets $\mathcal{D}_{k\in \mathcal{K}}$ are preserved at the users and not transmitted to the BS. Hence, FL-based model training is performed at the user side as
		\begin{align}
		\label{lossFL}
		\minimize_{\boldsymbol{\theta}}  & \hspace{10pt} 
		\bar{\mathcal{L}}(\boldsymbol{\theta})  = \frac{1}{K}\sum_{k=1}^{K} \mathcal{L}_k(\boldsymbol{\theta})  \nonumber\\
		\subjectto & \hspace{10pt}f(\mathcal{X}_k^{(i)}|\boldsymbol{\theta}) = \mathcal{Y}_k^{(i)}, i = 1,\dots, \textsf{D}_k, k\in \mathcal{K},
		\end{align}
		where $\mathcal{L}_k(\boldsymbol{\theta}) =  \frac{1}{\textsf{D}_k}\sum_{i=1}^{\textsf{D}_k}\| f( \mathcal{X}_k^{(i)}|\boldsymbol{\theta}) - \mathcal{Y}_k^{(i)}  \|_\mathcal{F}^2$. Notice that the FL-based model training in (\ref{lossFL}) is solved at the user while the CL problem in (\ref{lossML}) is handled at the BS.}  To efficiently solve (\ref{lossFL}) and (\ref{lossML}), gradient descent (GD) is employed and the problems are solved iteratively. In CL, the gradient is computed over the whole dataset as $	 \mathbf{g}(\boldsymbol{\theta}_t) = \nabla \mathcal{L}(\boldsymbol{\theta}_t)$	and the parameter update is performed as 
	\begin{align}
	\label{eq:UpdateCL}
	\boldsymbol{\theta}_{t+1} = \boldsymbol{\theta}_t - \eta  {\mathbf{g}}(\boldsymbol{\theta}_t),
	\end{align}
	where $\eta$ is the learning rate.
	
	In FL, each user computes the gradients individually as $	\mathbf{g}_k(\boldsymbol{\theta}_t) = \nabla \mathcal{L}_k(\boldsymbol{\theta}_t)$ to solve (\ref{lossFL}), then sends them to the BS, where the model parameters are updated as
	\begin{align}
	\label{eq:UpdateAtBSNoiseFree}
	\boldsymbol{\theta}_{t+1} = \boldsymbol{\theta}_t - \eta  \frac{1}{K} \sum_{k=1}^{K} {\mathbf{g}}_k(\boldsymbol{\theta}_t).
	\end{align}

	\textcolor{black}{The transmission of gradients to the BS provides more energy-efficiency than directly transmitting the model parameters as in the \texttt{FedAvg} algorithm~\cite{fl_By_Google}. The main reason is that gradients include only the model updates obtained from the GD algorithm, whereas model transmission includes already known data from the previous iteration. Hence, model transmission wastes a significant amount of transmit power from all the users~\cite{elbir2020FL_HB,FL_Gunduz,FL_QSGD}.}
	
	The gradients $\mathbf{g}_{k\in \mathcal{K}}(\boldsymbol{\theta}_t)$ are sent to the BS via wireless channel, which causes corruptions during transmission. {\color{black}Therefore, the corrupted model parameters and gradients at the $t$th iteration are given as~\cite{FL_gunduz_fading,robustFL}
		\begin{align}
		\tilde{\boldsymbol{\theta}}_t &= {\boldsymbol{\theta}}_t + \Delta{\boldsymbol{\theta}}_t,\label{gradientNoisya} \\
		\mathbf{g}_k(\tilde{\boldsymbol{\theta}}_t) &= \mathbf{g}_k({\boldsymbol{\theta}}_t) + \Delta\mathbf{g}_k({\boldsymbol{\theta}}_t), \label{gradientNoisyb} \\
		\tilde{\mathbf{g}}_k(\tilde{\boldsymbol{\theta}}_t) &= \mathbf{g}_k(\tilde{\boldsymbol{\theta}}_t) + \Delta\mathbf{g}_k(\tilde{\boldsymbol{\theta}}_t),\label{gradientNoisyc}
		\end{align}
		where $\tilde{\boldsymbol{\theta}}_t$ represents the noisy model parameters captured at the users, $\mathbf{g}_k(\tilde{\boldsymbol{\theta}}_t)$ is the gradient vector computed at the user based on $\tilde{\boldsymbol{\theta}}_t$ and $\tilde{\mathbf{g}}_k(\tilde{\boldsymbol{\theta}}_t)$ denotes the noisy gradient vector received at the BS. $\Delta\boldsymbol{\theta}_t$, $\Delta\mathbf{g}_k({\boldsymbol{\theta}}_t)$ and $\Delta\mathbf{g}_k(\tilde{\boldsymbol{\theta}}_t)$ represent the noise terms added onto $\boldsymbol{\theta}_t$, $\mathbf{g}_k({\boldsymbol{\theta}}_t)$ and $\mathbf{g}_k(\tilde{\boldsymbol{\theta}}_t)$, respectively.  Then, the model update rule can be given by
		\begin{align}
		\tilde{\boldsymbol{\theta}}_{t+1} = \tilde{\boldsymbol{\theta}}_t - \eta \frac{1}{K}\sum_{k=1}^K \tilde{\mathbf{g}}_k(\tilde{\boldsymbol{\theta}}_t) ,
		\end{align}
		which can be rewritten as
		\begin{align}
		&\tilde{\boldsymbol{\theta}}_{t+1}= \big[\boldsymbol{\theta}_t + \Delta{\boldsymbol{\theta}}_t \big]   -\eta \sum_{k=1}^K\frac{\big[{\mathbf{g}}_k({\boldsymbol{\theta}}_t) + \Delta\mathbf{g}_k({\boldsymbol{\theta}}_t) + \Delta\mathbf{g}_k(\tilde{\boldsymbol{\theta}}_t) \big]}{K} \nonumber \\
		&= \underbrace{\boldsymbol{\theta}_t - \eta \sum_{k=1}^K \frac{\mathbf{g}_k(\boldsymbol{\theta}_t)}{K}}_{\boldsymbol{\theta}_{t+1}}
		+ \underbrace{\Delta{\boldsymbol{\theta}}_t - \eta\sum_{k=1}^K  \frac{\big[\Delta\mathbf{g}_k({\boldsymbol{\theta}}_t) + \Delta\mathbf{g}_k(\tilde{\boldsymbol{\theta}}_t)  \big]}{K}     }_{ \Delta } \nonumber \\
		&=  \boldsymbol{\theta}_{t+1} + \Delta,
		\end{align}
		where $\Delta$ corresponds to the overall noise term added onto $\boldsymbol{\theta}_{t+1} = \boldsymbol{\theta}_t - \eta \frac{1}{K}\sum_{k=1}^K \mathbf{g}_k(\boldsymbol{\theta}_t)$. Now, let us consider the statistics of $\Delta$. Without loss of generality, the noise terms due to wireless transmission in (\ref{gradientNoisya}) and (\ref{gradientNoisyc}), i.e., $\Delta{\boldsymbol{\theta}}_t$ and $\Delta\mathbf{g}_k(\tilde{\boldsymbol{\theta}}_t)$, can be modeled as AWGN with variances $\sigma_{\boldsymbol{\theta}}^2$ and $\tilde{\sigma}_k^2$, respectively~\cite{fl_convergenceOnNIIDData,robustFL}. Furthermore, we define $\Delta\mathbf{g}_k({\boldsymbol{\theta}}_t)$ in  (\ref{gradientNoisyb}) as AWGN with variance ${\sigma}_k^2$ due to the linearity of gradient and the NN layers\footnote{Many of the NN layers, such as convolutional, fully connected, normalization and dropout layers, perform linear operations, whereas pooling and ReLU layers are non-linear~\cite{deepLearningScience,robustFL,fl_convergenceOnNIIDData}.}. Hence, the overall noise term $\Delta$ can be viewed as an AWGN with variance $\sigma_\Delta^2 = \sigma_{\boldsymbol{\theta}}^2 + \eta \frac{\sum_{k=1}^{K}(\tilde{\sigma}_k^2+ {\sigma}_k^2)}{K}$.
		
		In order to solve (\ref{lossFL}) effectively in the presence of noisy model parameters, we define a regularized loss function $\tilde{\mathcal{L}}_k(\boldsymbol{\theta})$ as
		\begin{align}
		\label{lossFLRegularized}
		\tilde{\mathcal{L}}_k(\boldsymbol{\theta}) = \mathcal{L}_k(\boldsymbol{\theta}) + \sigma_\Delta^2 ||\mathbf{g}_k(\boldsymbol{\theta}) ||^2,
		\end{align}
		which is widely used in stochastic optimization~\cite{stochasticOpt}. (\ref{lossFLRegularized})  can be obtained via  first order Taylor  expansion of the expectation-based loss $\mathbb{E}\{|| \mathcal{L}_k(\boldsymbol{\theta} +\Delta )||^2 \}$, which can be approximately written as 
		\begin{align}
		\mathbb{E}\{|| \mathcal{L}_k(\boldsymbol{\theta} + &\Delta )||^2 \}\approx \mathbb{E}\{|| \mathcal{L}_k(\boldsymbol{\theta} ) + {\Delta}\nabla\mathcal{L}_k(\boldsymbol{\theta} )     ||^2 \}, \nonumber \\
		&\approx \mathbb{E}\{|| \mathcal{L}_k(\boldsymbol{\theta} )||^2 \}  + \mathbb{E}\{||\Delta ||^2 \} \mathbb{E}\{|| \nabla\mathcal{L}_k(\boldsymbol{\theta} )     ||^2 \}, \nonumber \\
		& \approx \mathbb{E}\{|| \mathcal{L}_k(\boldsymbol{\theta} )||^2 \} +   {\sigma}_{\Delta}^2 ||\mathbf{g}(\boldsymbol{\theta}) ||^2,
		\end{align}
		where the first term corresponds to the minimization of the loss function with perfect estimation and the second term is the additional cost due to noise~\cite{stochasticOpt,robustFL}. Using  (\ref{lossFLRegularized}), the regularized version of FL-based training problem in (\ref{lossFL}) is given by
		\begin{align}
		\minimize_{\boldsymbol{\theta}}& \hspace{10pt} \bar{\mathcal{L}}(\boldsymbol{\theta}) = \frac{1}{K} \sum_{k=1}^{K} \tilde{\mathcal{L}}_k(\boldsymbol{\theta}) \nonumber \\
		\subjectto & \hspace{10pt}f(\mathcal{X}_k^{(i)}|\boldsymbol{\theta}) = \mathcal{Y}_k^{(i)}, i = 1,\dots, \textsf{D}_k, k\in \mathcal{K},
		\end{align} 
		which can be effectively solved via GD in the presence of noisy model updates as
		\begin{align}
		\tilde{\boldsymbol{\theta}}_{t+1} = \tilde{\boldsymbol{\theta}}_t - \eta \nabla \bar{\mathcal{L}}(\tilde{\boldsymbol{\theta}}),
		\end{align}
		where $\nabla\bar{\mathcal{L}}(\tilde{\boldsymbol{\theta}}) =\frac{1}{K} \sum_{k=1}^K \bar{\mathbf{g}}_k(\tilde{\boldsymbol{\theta}}_t)$ and  $\bar{\mathbf{g}}_k(\tilde{\boldsymbol{\theta}}_t) = \nabla \tilde{\mathcal{L}}_k(\tilde{\boldsymbol{\theta}}_t) =  \nabla \big[\mathcal{L}_k(\tilde{\boldsymbol{\theta}}_t) + \sigma_\Delta^2 ||\mathbf{g}_k(\tilde{\boldsymbol{\theta}}_t) ||^2\big]$. 
		
		Due to the effect of noisy gradient transmission, $\bar{\mathcal{L}}(\boldsymbol{\theta})$ converges slower than ${\mathcal{L}}(\boldsymbol{\theta})$. In the following theorem, we prove the convergence of  $\bar{\mathcal{L}}(\boldsymbol{\theta})$. While the convergence of the regularized loss function was studied in different FL works~\cite{fl_convergenceOnNIIDData,robustFL}, they consider model transmission, whereas in this work we investigate the gradient transmission approach. The convergence analysis is also different from the previous  gradient transmission-based works, e.g.,~\cite{FL_Gunduz,FL_gunduz_fading}, which are based on the sparsity assumption of the gradient vector, which may not be always satisfied.  
		
		\textit{Theorem 1:} Let $\boldsymbol{\theta}_{0}$ and $\boldsymbol{\theta}_{\star}$ be the initial and optimal model parameters, respectively. Then, the FL-based model training converges with the convergence rate $\mathcal{O}(1/t)$ as 
		\begin{align}
		\bar{\mathcal{L}}(\boldsymbol{\theta}_{t}) - \bar{\mathcal{L}}(\boldsymbol{\theta}_\star) \leq ||\boldsymbol{\theta}_{0} - \boldsymbol{\theta}_\star  ||^2  \frac{1}{2\eta  }  \frac{ 1}{t},
		\end{align}
		with the learning rate $\eta \leq \frac{1}{(1+{\sigma}_\Delta ^2) \beta }$ for some $\beta \geq 0$.
		
		\textit{Proof:} See Appendix~\ref{appendix1}.\qed 
		
	}

	\textcolor{black}{In practice, the convergence of the learning model is subject to the wireless factors, such as the SNR of the transmitted/received model updates. In particular, the convergence becomes slower due to the packet errors during training~\cite{fl_convergence_packetErrorRate}. Furthermore, the channel statistics change in each communication round, which entails CSI acquisition for each round. While some of the recent works assume that a single communication round between the server and the clients takes a single channel coherence time~\cite{elbir2020FL_HB,FL_Gunduz,FL_gunduz_fading}, in~\cite{fl_in_singleCoherenceInterval} FL-based training is completed in a single long-coherence time, which is approximately composed of 40 small-scale fading channel coherence intervals~\cite{fl_in_singleCoherenceInterval}.    }

	\subsection{FL for Channel Estimation in Massive MIMO}
	\label{sec:FLmMIMO}
	Here, we discuss how the input and output of \textsf{ChannelNet} are determined for massive MIMO scenario.	
	
	The input of \textsf{ChannelNet} is the set of received pilot signals at the preamble stage. 	Consider the downlink received signal model in (\ref{eq:ReceivedMIMO}) and assume that the BS activates only a single RF chain, one at a time. Let $\overline{\mathbf{f}}_u[m]\in\mathbb{C}^{N_\mathrm{BS}}$ be the resulting beamformer vector and pilot signals are $\overline{{s}}_u[m]$, where  $u = 1,\dots,M_\mathrm{BS}$ and $m\in \mathcal{M}$. At the receiver side, each user activates the RF chain for $M_\mathrm{MS}$ times and applies the beamformer vector  $\overline{\mathbf{w}}_v[m]$, $v = 1,\dots, M_\mathrm{MS}$  to process the received pilots \cite{deepCNN_ChannelEstimation}.  Hence, the total channel use in the channel acquisition process is $M_\mathrm{BS}\lceil \frac{M_\mathrm{MS}}{N_\mathrm{RF}}\rceil$. Therefore, the received pilot signal at the $k$th user becomes
	\begin{align}
	\label{receivedSignalPilot}
	\mathbf{\overline{Y}}_k[m] = \overline{\mathbf{W}}^\textsf{H}[m] \mathbf{H}_k[m] \overline{\mathbf{F}}[m]\overline{\mathbf{S}}[m] + \widetilde{\mathbf{N}}_k[m],
	\end{align}
	where $\overline{\mathbf{F}}[m] = [\overline{\mathbf{f}}_1[m],\overline{\mathbf{f}}_2[m],\dots,\overline{\mathbf{f}}_{M_\mathrm{BS}}[m]]$ and $\overline{\mathbf{W}}[m] = [\overline{\mathbf{w}}_1[m],\overline{\mathbf{w}}_2[m],\dots,\overline{\mathbf{w}}_{M_\mathrm{MS}}[m]]$ are $N_\mathrm{BS}\times M_\mathrm{BS}$ and $N_\mathrm{MS}\times M_\mathrm{MS}$ beamformer matrices, respectively.  $\overline{\mathbf{S}}[m] = \mathrm{diag}\{ \overline{s}_1[m],\dots,\overline{s}_{M_\mathrm{BS}}[m]\}$ denotes pilot signals and $\widetilde{\mathbf{N}}_k[m]= \overline{\mathbf{W}}^\textsf{H} \overline{\mathbf{N}}_k[m]$ is the effective noise matrix, where $\overline{\mathbf{N}}_k[m] \sim \mathcal{N}(0, \sigma^2\mathbf{I}_{M_\mathrm{MS}})$. 
	{\color{black}Without loss of generality, we assume that $\overline{\mathbf{F}}[m] = \overline{\mathbf{F}}$ and $\overline{\mathbf{W}}[m] = \overline{\mathbf{W}}$ and $\overline{\mathbf{S}}[m] = \mathbf{I}_{M_\mathrm{BS}}$ $\forall m\in\mathcal{M}$. Then, the received signal (\ref{receivedSignalPilot}) becomes
		\begin{align}
		\label{receivedSignalPilotMod}
		\mathbf{\overline{Y}}_k[m] = \overline{\mathbf{W}}^\textsf{H} \mathbf{H}_k[m] \overline{\mathbf{F}} + \widetilde{\mathbf{N}}_k[m].
		\end{align}
		Using $\mathbf{\overline{Y}}_k[m]$, we define the input of \textsf{ChannelNet} $\mathbf{G}_k[m]$ as
		\begin{align}
		\label{Gm}
		\mathbf{G}_k[m] = \mathbf{T}_\mathrm{MS} \overline{\mathbf{Y}}_k[m]\mathbf{T}_\mathrm{BS},
		\end{align}
		where $\mathbf{T}_\mathrm{MS} = 
		\begin{dcases}
		\overline{\mathbf{W}},& M_\mathrm{MS} < N_\mathrm{MS} \\
		(\overline{\mathbf{W}}\overline{\mathbf{W}}^\textsf{H})^{-1}\overline{\mathbf{W}}, & M_\mathrm{MS} \geq N_\mathrm{MS},
		\end{dcases}$ and $	\mathbf{T}_\mathrm{BS}  = 
		\begin{dcases} 
		\overline{\mathbf{F}}^\textsf{H},& M_\mathrm{BS} < N_\mathrm{BS} \\
		\overline{\mathbf{F}}^\textsf{H}(\overline{\mathbf{F}}\overline{\mathbf{F}}^\textsf{H})^{-1}, & M_\mathrm{BS} \geq N_\mathrm{BS}.
		\end{dcases}$
		Here,  $\mathbf{T}_\mathrm{BS}$ and $\mathbf{T}_\mathrm{MS}$ clear the effect of unitary matrices $\overline{\mathbf{F}}$ and $\overline{\mathbf{W}}$ in (\ref{receivedSignalPilotMod}), respectively. Since \textsf{ChannelNet} accepts real-valued data, we construct the final form of the input $\mathcal{X}_k$ as  three ``channel'' tensors. Thus, the first and second ``channel'' of $\mathcal{X}_k$ are the real and imaginary parts of $\mathbf{G}_k[m]$, i.e., $[\mathcal{X}_k]_1 = \operatorname{Re}\{\mathbf{G}_k[m]\}$ and $[\mathcal{X}_k]_2 = \operatorname{Im}\{\mathbf{G}_k[m]\}$, respectively. Finally, the third ``channel'' is given by $[\mathcal{X}_k]_3 = \angle\{\mathbf{G}_k[m]\}$.  We note here that the use of three ``channel'' input (e.g., real, imaginary and angle information of $\mathbf{G}_k[m]$) provides better feature representation~\cite{elbirDL_COMML,elbir_LIS,elbir2020withoutCSI}. As a result, the size of the input data is $N_\mathrm{MS}\times N_\mathrm{BS}\times 3$.	
		
		The output of \textsf{ChannelNet} is given by a  $2N_\mathrm{BS}N_\mathrm{MS}\times 1$ real-valued vector as 
		\begin{align}
		\mathcal{Y}_k = \left[\mathrm{vec}\{\operatorname{Re}\{\mathbf{H}_k[m]\}\}^\textsf{T},\mathrm{vec}\{\operatorname{Im}\{\mathbf{H}_k[m]\}\}^\textsf{T}\right]^\textsf{T}.
		\end{align}
		
		As a result, \textsf{ChannelNet} maps the received pilot signals $\mathbf{G}_k[m]$ to the channel matrix $\mathbf{H}_k[m]$.
		
		

		\subsection{FL for Channel Estimation in RIS-Assisted Massive MIMO}
		\label{sec:FL_RIS}
		In this part, we examine the channel estimation problem in RIS-assisted massive MIMO, which is shown in Fig.~\ref{fig_diagRIS}. First, we present the received signal model including both direct (BS-user) and cascaded (BS-RIS-user) channels\textcolor{black}{\footnote{\textcolor{black}{Channel estimation is required to design the passive beamformer weights. Although the BS-user, BS-RIS and RIS-user channels can be estimated separately~\cite{irs_CE_SeparateChannels}, the estimation of the direct and the cascaded channels is sufficient for beamformer design~\cite{irs_beamformingWithCascaded1,irs_beamformingWithCascaded2}.}}}. Then, we show how input-output pairs of \textsf{ChannelNet} are obtained for RIS-assisted scenario.

		%
		\begin{figure}[h]
			\centering
			{\includegraphics[draft=false,width=.8\columnwidth]{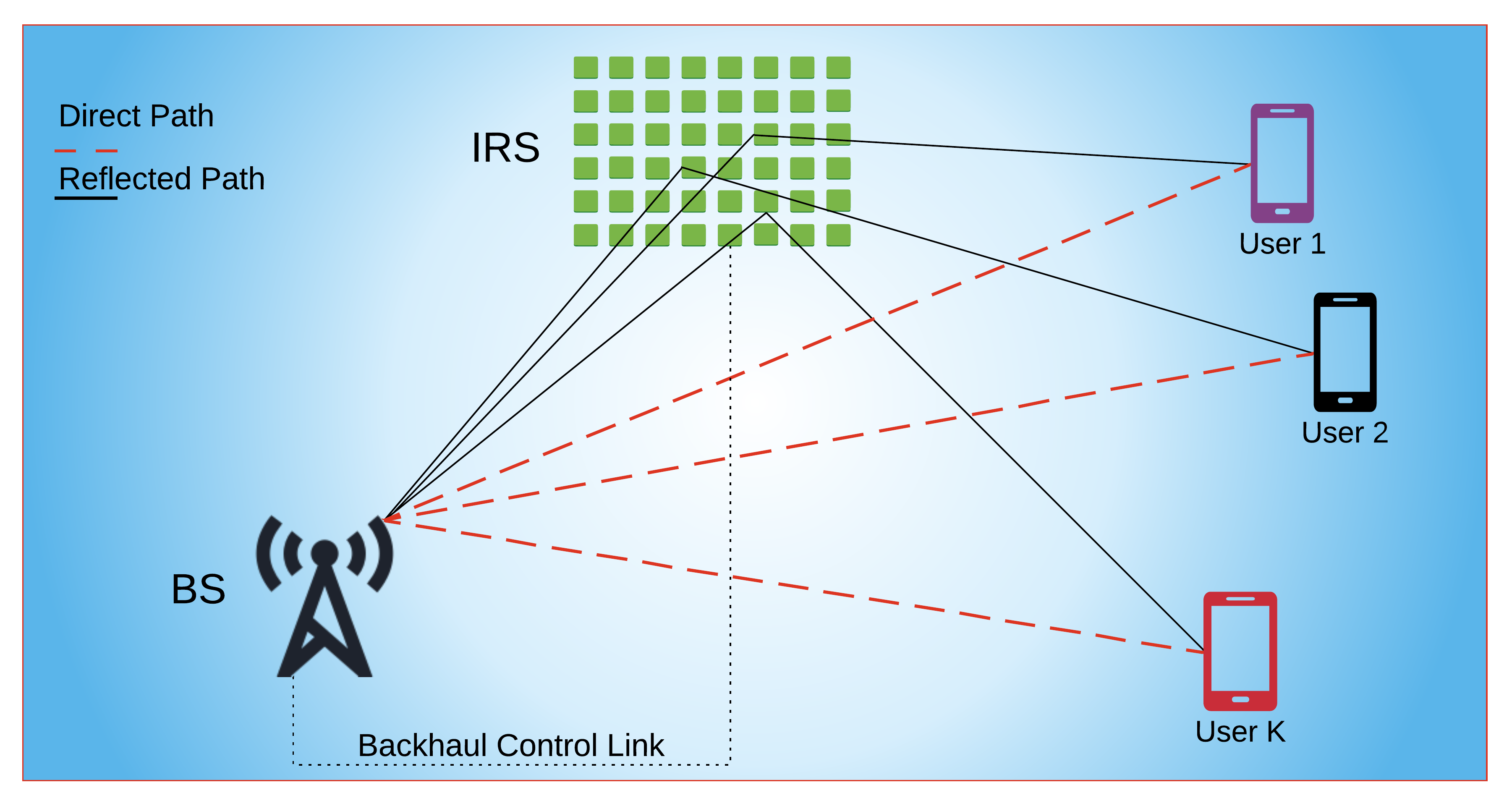} } 
			\caption{RIS-assisted mm-Wave massive MIMO scenario.
			}
			\label{fig_diagRIS}
		\end{figure}

		We consider the downlink channel estimation, where the BS has $N_\mathrm{BS}$ antennas to serve $K$ single-antenna users with the assistance of RIS, which is composed of $N_\mathrm{RIS}$ reflective elements, as shown in Fig.~\ref{fig_diagRIS}. The incoming signal from the BS is reflected from the RIS, where each RIS element introduces a phase shift $\varphi_n$, for $n=1,\dots, N_\mathrm{RIS}$. This phase shift can be adjusted through the PIN (positive-intrinsic-negative) diodes, which are controlled by the RIS-controller connected to the BS over the backhaul link. As a result, RIS allows the users receive the signal transmitted from the BS when they are distant from the BS or there is a blockage among them. 	Let $\overline{\mathbf{S}}_\mathrm{RIS}\in \mathbb{C}^{N_\mathrm{BS}\times M_\mathrm{BS}}$, $(N_\mathrm{BS} \leq M_\mathrm{BS})$ be the pilot signals transmitted from the BS, then the received signal at the $k$th user becomes
		\begin{align}
		\label{receivedRIS1}
		\mathbf{y}_k = (\mathbf{h}_{\mathrm{B},k}^\textsf{H} + \boldsymbol{\psi}^\textsf{H}\mathbf{V}_k^\textsf{H}) \overline{\mathbf{S}}_\mathrm{RIS} + \mathbf{n}_k,
		\end{align}
		where $\mathbf{y}_k = [y_{1,k},\dots, y_{M_\mathrm{BS},k}]$ and $\mathbf{n}_k = [n_{1,k},\dots, n_{M_\mathrm{BS},k}]$ are $1\times M_\mathrm{BS}$ row vectors and $\mathbf{h}_{\mathrm{B},k}\in \mathbb{C}^{N_\mathrm{BS}}$ represents the channel for the communication link between the BS and the $k$th user. 	 $\boldsymbol{\psi} = [\psi_1, \dots, \psi_{N_\mathrm{RIS}}]^\textsf{T}\in \mathbb{C}^{N_\mathrm{RIS}}$ is the reflecting beamformer vector, whose $n$th entry is $\psi_n = a_n e^{j \varphi_n} $, where $a_n\in \{0,1\}$ denotes the on/off stage of the $n$th element of the RIS and $\varphi_n \in [0, 2\pi]$ is the phase shift introduced by the RIS. In practice, the RIS elements cannot be perfectly turned on/off, hence, they can be modeled as $	a_n=\left\{\begin{array}{cc}
		1 - \epsilon_1 & \mathrm{ON}\\
		0 + \epsilon_0 & \mathrm{OFF}
		\end{array}\right.,$
		for small $\epsilon_1,\epsilon_0 \geq0$, which represent the insertion loss of the reflecting elements~\cite{elbir_LIS}. In (\ref{receivedRIS1}), $\mathbf{V}_k\in \mathbb{C}^{N_\mathrm{BS}\times N_\mathrm{RIS}}$ denotes the cascaded channel for the BS-RIS-user link and it can be defined in terms of the channel between BS-RIS and RIS-user as 
		\begin{align}
		\label{cascadedChannel1}
		\mathbf{V}_k = \mathbf{H}_\mathrm{B} \boldsymbol{\Lambda}_k,
		\end{align}
		where $\mathbf{H}_{\mathrm{B}}\in \mathbb{C}^{N_\mathrm{BS} \times N_\mathrm{RIS}}$ is the channel between the BS and the RIS and it can be defined similar to (\ref{eq:delaydChannelModel}) as
		\begin{align}
		\mathbf{H}_{\mathrm{B}} = \sqrt{\frac{N_\mathrm{BS}N_\mathrm{RIS} }{L_\mathrm{RIS}  }} \sum_{l=1}^{ L_\mathrm{RIS}} \alpha_{l}^{\mathrm{RIS}} \mathbf{a}_\mathrm{BS}( \phi_{l}^{\mathrm{BS}})\mathbf{a}_\mathrm{RIS}( {\phi}_{l}^{\mathrm{RIS}})^\textsf{H},
		\end{align}
		where $L_\mathrm{RIS}$ and $\alpha_{l}^{\mathrm{RIS}}$ are the number of received paths and the complex gain respectively. $\mathbf{a}_\mathrm{BS}( \phi_{l}^{\mathrm{BS}})\in \mathbb{C}^{N_\mathrm{BS}}$ and $\mathbf{a}_\mathrm{RIS}( {\phi}_l^{\mathrm{RIS}})\in \mathbb{C}^{N_\mathrm{RIS}}$ are the steering vectors corresponding to the BS and RIS with the AoA and AoD angles of  $\phi_{l}^{\mathrm{BS}},\phi_{l}^{\mathrm{RIS}}$, respectively. In (\ref{cascadedChannel1}), $\boldsymbol{\Lambda}_k =  \mathrm{diag}\{ \mathbf{h}_{\mathrm{S},k}\}$ and 
		$\mathbf{h}_{\mathrm{S},k}\in \mathbb{C}^{N_\mathrm{RIS}}$ represents the channel between the RIS and the $k$th user. $\mathbf{h}_{\mathrm{S},k}$ and $\mathbf{h}_{\mathrm{B},k}$ have similar structure and they can be defined as follows
		\begin{align}
		\mathbf{h}_{\mathrm{B},k} &= \sqrt{\frac{N_\mathrm{BS} }{L_\mathrm{B}  }} \sum_{l=1}^{ L_\mathrm{B}} \alpha_{k,l}^{\mathrm{B}} \mathbf{a}_\mathrm{BS}( \phi_{k,l}^{\mathrm{B}}), \\
		\mathbf{h}_{\mathrm{S},k} & = \sqrt{\frac{N_\mathrm{RIS} }{L_\mathrm{S}  }} \sum_{l=1}^{ L_\mathrm{S}} \alpha_{k,l}^{\mathrm{S}} \mathbf{a}_\mathrm{RIS}( \phi_{k,l}^{\mathrm{S}}),
		\end{align}
		where $L_\mathrm{B}$, $\alpha_{k,l}^{\mathrm{B}}$ and $\mathbf{a}_\mathrm{BS}(\phi_{k,l}^{\mathrm{B}})$ ($L_\mathrm{S}$, $\alpha_{k,l}^{\mathrm{S}}$, $\mathbf{a}_\mathrm{RIS}(\phi_{k,l}^{\mathrm{B}})$) are the number of paths, complex gain and the steering vector for the BS-user (RIS-user) communication link, respectively.

		In order to estimate the direct channel $\mathbf{h}_{\mathrm{B},k}$, we assume that all the RIS elements are turned off, i.e., $a_n = 0$ for $n = 1,\dots, N_\mathrm{RIS}$. Then, the $1\times M_\mathrm{BS}$ received signal at the $k$th user becomes
		\begin{align}
		\label{esthdirect}
		\mathbf{y}_{\mathrm{B},k} = \mathbf{h}_{\mathrm{B},k}^\textsf{H}\overline{\mathbf{S}}_\mathrm{RIS} + \mathbf{n}_{\mathrm{B},k}.
		\end{align}
		Then, the direct channel between BS-user $\mathbf{h}_{\mathrm{B},k}$ can be estimated from the received pilot signal $\mathbf{y}_{\mathrm{B},k}$ {\color{black}via LS and MMSE estimators as $	\mathbf{h}_{\mathrm{B},k}^\mathrm{LS} = \bigg(\mathbf{y}_{\mathrm{B},k}\overline{\mathbf{S}}_\mathrm{RIS}^\dagger\bigg)^\textsf{H},$ and $\mathbf{h}_{\mathrm{B},k}^\mathrm{MMSE} = \mathbf{h}_{\mathrm{B},k}^\mathrm{LS} \mathbf{R}_{\mathrm{B},k} \bigg(\mathbf{R}_{\mathrm{B},k} + \frac{1}{\sigma^2} \overline{\mathbf{S}}_\mathrm{RIS}\overline{\mathbf{S}}_\mathrm{RIS}^\textsf{H}  \bigg)^{-1},$
%
%
			where $\mathbf{R}_{\mathrm{B},k} = \mathbb{E}\{ \mathbf{h}_{\mathrm{B},k} \mathbf{h}_{\mathrm{B},k}^\textsf{H}\}$~\cite{deepCNN_ChannelEstimation}}.
		
		Next, we consider the cascaded channel estimation. We assume that each RIS element is turned on one by one while all the other elements are turned off. This is done by the BS requesting the RIS via a micro-controller device in the backhaul link so that a single RIS element is turned on at a time. Then, the reflecting beamformer vector at the $n$th frame becomes $\boldsymbol{\psi}^{(n)} = [0,\dots, 0, \psi_n,0,\dots, 0]^\textsf{T}$, where $a_n = \{0: \tilde{n} = 1,\dots N_\mathrm{RIS}, \tilde{n} \neq n \}$ and the received signal is given by
		\begin{align}
		\label{esthcascaded}
		\mathbf{y}_{\mathrm{C},k}^{(n)} = (\mathbf{h}_{\mathrm{B},k}^\textsf{H} +{\mathbf{v}_{k}^{(n)}}^\textsf{H} ) \overline{\mathbf{S}}_\mathrm{RIS} + \mathbf{n}_k,
		\end{align}
		where $\mathbf{v}_k^{(n)} \in \mathbb{C}^{N_\mathrm{BS}}$ is the $n$th column of $\mathbf{V}_k$, i.e., $\mathbf{v}_k^{(n)} = \mathbf{V}_k \boldsymbol{\psi}^{(n)}$, where  $\psi_n = 1$. Using the estimate of $\mathbf{h}_{\mathrm{B},k}$ from (\ref{esthdirect}), (\ref{esthcascaded}) can be solved for $\mathbf{v}_k^{(n)}$, $n = 1,\dots, N_\mathrm{RIS}$, and the cascaded channel $\mathbf{V}_k$ can be estimated. Then, the  received data for $n = 1\dots, N_\mathrm{RIS}$ can be given by $\mathbf{Y}_{\mathrm{C},k}\in \mathbb{C}^{N_\mathrm{RIS}\times M_\mathrm{BS}}$ as $\mathbf{Y}_{\mathrm{C},k} =\left[\begin{array}{c}
		\mathbf{y}_{\mathrm{C},k}^{(n)}\\
		\vdots \\
		\mathbf{y}_{\mathrm{C},k}^{(N_\mathrm{RIS})}
		\end{array}\right]$. In order to train \textsf{ChannelNet} for RIS-assisted massive MIMO scenario, we select  the input-output data pair as $\{\mathbf{y}_{\mathrm{B},k},\mathbf{h}_{\mathrm{B},k} \}$ and $\{\mathbf{Y}_{\mathrm{C},k},\mathbf{V}_{k} \}$ for direct and cascaded channels respectively. To jointly learn both channels, a single input is constructed to train a single NN as  $ \boldsymbol{\Upsilon}_k = \left[\begin{array}{c}
		\mathbf{y}_{\mathrm{B},k}\\
		\mathbf{Y}_{\mathrm{C},k}
		\end{array}\right] \in \mathbb{C}^{(N_\mathrm{RIS} + 1)\times M_\mathrm{BS}}$. Following the same strategy in the previous scenario,  the three ``channel" of the input data can be constructed as $[{\mathcal{X}}_k]_1 = \operatorname{Re}\{ \boldsymbol{\Upsilon}_k\}$ and $[{\mathcal{X}}_k]_2 = \operatorname{Im}\{ \boldsymbol{\Upsilon}_k\}$, $[{\mathcal{X}}_k]_3 = \angle\{ \boldsymbol{\Upsilon}_k\}$, respectively. We can define the output data as $\boldsymbol{\Sigma}_k = \left[\mathbf{h}_{\mathrm{B},k}, \mathbf{V}_{k}\right] \in \mathbb{C}^{N_\mathrm{BS}\times (N_\mathrm{RIS}+1)}$, hence, the output label can be given by a $2N_\mathrm{BS}(N_\mathrm{RIS}+1)\times 1$ real-valued vector as
		\begin{align}
		{\mathcal{Y}}_k = \left[ \mathrm{vec}\{\operatorname{Re} \{\boldsymbol{\Sigma}_k\}\}^\textsf{T},  \mathrm{vec}\{\operatorname{Im} \{\boldsymbol{\Sigma}_k\}\}^\textsf{T}   \right]^\textsf{T}.
		\end{align}
		
		Consequently, we have the sizes of ${\mathcal{X}}_k$ and ${\mathcal{Y}}_k$ are $(N_\mathrm{RIS} + 1)\times M_\mathrm{BS}\times 3$ and $2N_\mathrm{BS} (N_\mathrm{RIS}+1)\times 1$ respectively.

		\subsection{Neural Network Architecture \textcolor{black}{and Training}}
		\label{sec:NNarch}
		We design a single CNN, i.e., \textsf{ChannelNet} trained on two different datasets for both conventional and RIS-assisted massive MIMO applications. The proposed network architecture is a CNN with $10$ layers. The first layer is the input layer, which accepts the input data of size $N_\mathrm{MS}\times N_\mathrm{BS}\times 3$ and $(N_\mathrm{RIS} + 1)\times M_\mathrm{BS}\times 3$ for conventional and RIS-assisted massive MIMO scenario respectively.  The $\{2,4,6\}$th layers are the convolutional layers with $N_\mathrm{SF} = 128$ filters, each of which employs a $3\times 3$ kernel for 2-D spatial feature extraction. The $\{3,5,7\}$th layers are the normalization layers. The eighth layer is a fully connected layer with $N_\mathrm{FCL}=1024$ units, whose main purpose is to provide feature mapping. The ninth layer is a dropout layer with $\kappa=1/2$ probability. The dropout layer applies an $N_\mathrm{FCL}\times 1$ mask on the weights of the fully connected layer, whose elements are uniform randomly selected from $\{0,1\}$. As a result, at each iteration of FL training, randomly selected different set of weights in the fully connected layer is updated. Thus, the use of dropout layer reduces the size of $\boldsymbol{\theta}_t$ and $\mathbf{g}_k(\boldsymbol{\theta}_t)$, thereby, reducing model transmission overhead. Finally, the last layer is output regression layer, yielding the output channel estimate of size  $2N_\mathrm{MS}N_\mathrm{BS}\times 1$ and $2N_\mathrm{BS} (N_\mathrm{RIS}+1)\times 1$ for conventional and RIS-assisted massive MIMO applications respectively. 
		
		\textcolor{black}{During FL-based training, the collected datasets at the users are used to compute the model updates as in Section~\ref{sec:FLTraining} and transmitted to the BS. The collected model parameters at the BS are then aggregated as in (\ref{eq:UpdateAtBSNoiseFree}) and broadcast to the users for the next iteration. This process is conducted for $t=1,\dots, T$ communication rounds until convergence. }
		

		\begin{figure}[h]
			\centering
			{\includegraphics[draft=false,width=.9\columnwidth]{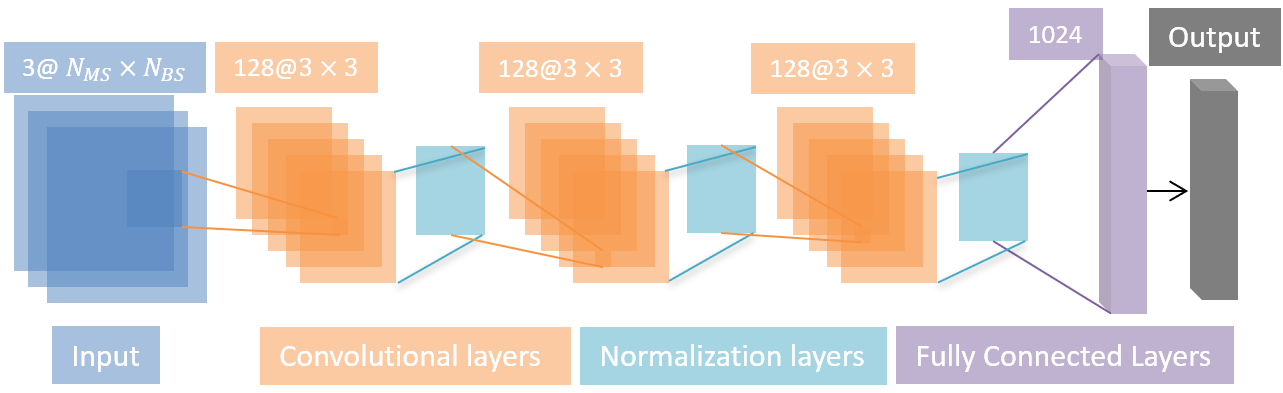} } 
			\caption{\color{black}The proposed CNN architecture for channel estimation. 
			}
			\label{fig_cnn}
		\end{figure}

		\section{Communication Overhead and Complexity}
		\label{sec:Complexity}
		\subsection{Communication Overhead}
		\label{sec:DataComp}
		Communication overhead can be defined as the size of the transmitted data during model training. Let $\mathcal{T}_\mathrm{FL}$ and $\mathcal{T}_\mathrm{CL}$ denote the communication overhead of FL and CL, respectively. Then, we can define $\mathcal{T}_\mathrm{CL}$ for both conventional and RIS-assisted scenario as
		\begin{align}
		\label{overheadCL}
		\mathcal{T}_\mathrm{CL}\hspace{-3pt} = \hspace{-3pt}\left\{\begin{array}{ll}
		\hspace{-3pt}\footnotesize (3N_\mathrm{MS} N_\mathrm{BS} + 2N_\mathrm{MS}N_\mathrm{BS}) \textsf{D}, & \hspace{-8pt}\footnotesize\mathrm{mMIMO}\\
		\hspace{-3pt}\footnotesize(3(N_\mathrm{RIS} + 1) M_\mathrm{BS}+ 2N_\mathrm{BS} (N_\mathrm{RIS} + 1)) \textsf{D}, & \hspace{-8pt}\footnotesize\mathrm{RIS}
		\end{array}\right. \hspace{-8pt},
		\end{align}
		which includes the number of symbols in the uplink transmission of the training dataset from the users to the BS. In contrast, the communication overhead of FL includes the transmission of $\mathbf{g}_k(\boldsymbol{\theta}_t)$ and $\boldsymbol{\theta}_t$ in uplink and downlink communication for $t = 1,\dots,T$, respectively. Hence, $\mathcal{T}_\mathrm{FL}$ is given by
		\begin{align}
		\label{overheadFL}
		\mathcal{T}_\mathrm{FL} = \left\{\begin{array}{ll}
		2PTK, & \mathrm{mMIMO}\\
		2PTK, & \mathrm{RIS}
		\end{array}\right..
		\end{align}
		We can see that the dominant terms in (\ref{overheadCL}) and (\ref{overheadFL}) are $\textsf{D}$ and $P$, which are the number of training data pairs and the number of NN parameters respectively.  While $\textsf{D}$ can be adjusted according to the amount of available data at the users, $P$ is usually unchanged during model training. 
		Here,  $P$ is computed as $P = \underbrace{N_\mathrm{CL}(CN_\mathrm{SF} W_x W_y)}_{\mathrm{Convolutional\hspace{1pt} Layers}} +   \underbrace{\kappa N_\mathrm{SF}  W_x W_yN_\mathrm{FCL} \footnotesize }_{\mathrm{Fully\hspace{1pt} Connected\hspace{1pt} Layers}},$
		where  $N_\mathrm{CL}=3$ is the number of convolutional layers and $C=3$ is the number of spatial ``channels''.  $W_x=W_y=3$ are the 2-D kernel sizes. As a result, we have $P=600,192$.  \textcolor{black}{Since the number of samples in the training dataset is usually larger than the number of model parameters, it is expected to have $\mathcal{T}_\mathrm{FL} < \mathcal{T}_\mathrm{CL}$~\cite{fl_By_Google,FL_Gunduz,elbir2020FL_HB} (see Fig.~\ref{fig_transmissionOverhead}). } 
		
		\begin{table}[ht]
			\caption{Convolutional Layers Settings}
			\label{tableConvLayerParameters}
			\centering
			\begin{tabular}{|c|c|c|c|c|c|c|}
				\hline
				$l$&$D_x^{(l)}$& $D_y^{(l)}$ &$W_x^{(l)}$  & $W_y^{(l)}$  &   $N_\mathrm{SF}^{(l-1)}$  & $N_\mathrm{SF}^{(l)}$          \\
				\hline
				2 &	$N_\mathrm{MS}$ & $N_\mathrm{BS}$ &3	 & 3&	3 &128 \\
				\hline
				4  &	$N_\mathrm{MS}$ &  $N_\mathrm{BS}$ &3	 &3 &	128 &128 \\
				\hline
				6 &	$N_\mathrm{MS}$ & $N_\mathrm{BS}$ 	& 3 &3 &128	 &128\\
				\hline 
			\end{tabular}
		\end{table}

		\begin{figure}[h]
			\centering
			{\includegraphics[draft=false,width=.9\columnwidth]{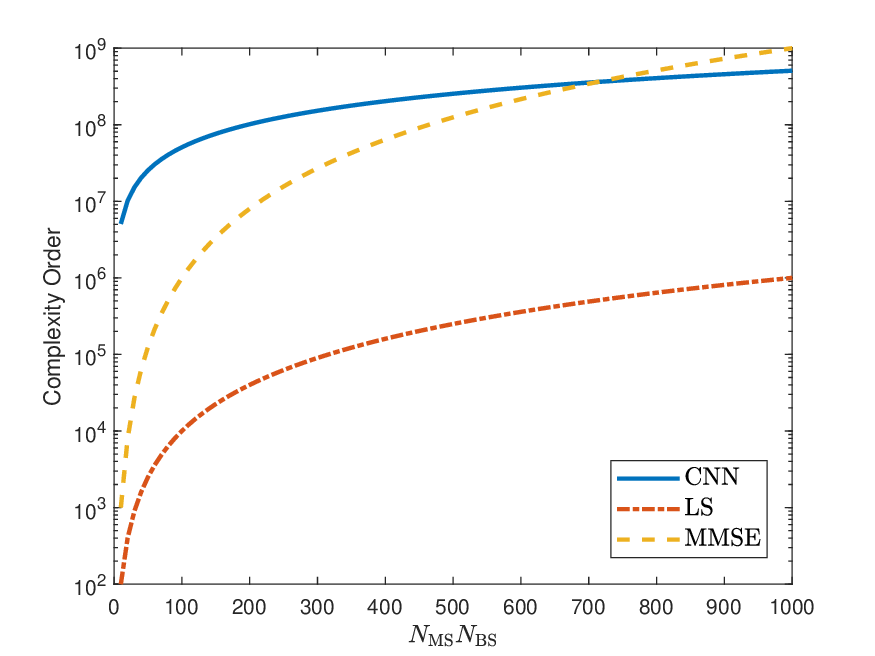} } 
			\caption{\color{black}Complexity order for CNN, MMSE and LS for channel estimation. 
			}
			\label{fig_Complexity}
		\end{figure}

		\begin{figure}[ht]
			\centering
			\subfloat[]{\includegraphics[draft=false,width=.5\columnwidth]{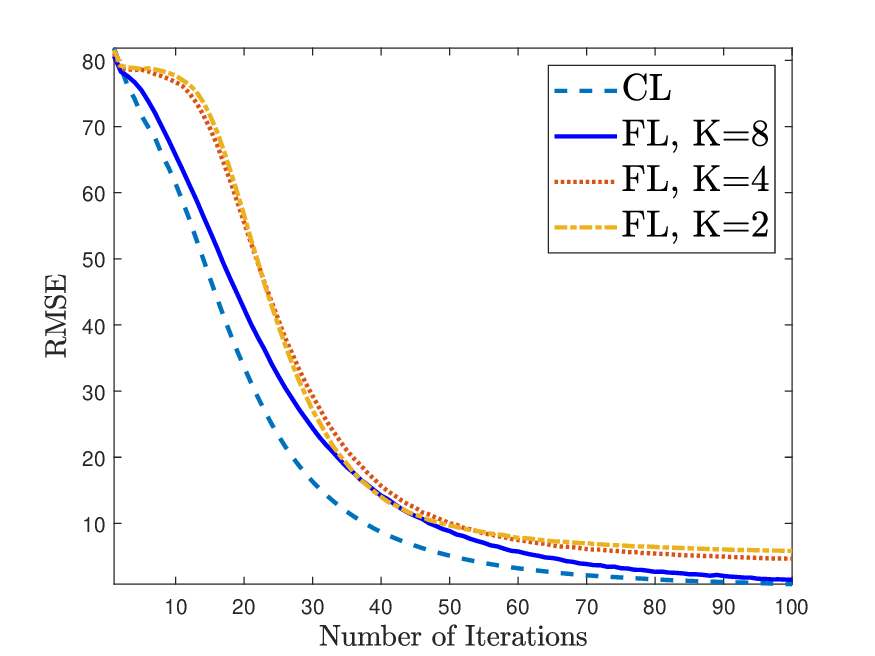}\label{fig_U_all} } 
			\subfloat[]{\includegraphics[draft=false,width=.5\columnwidth]{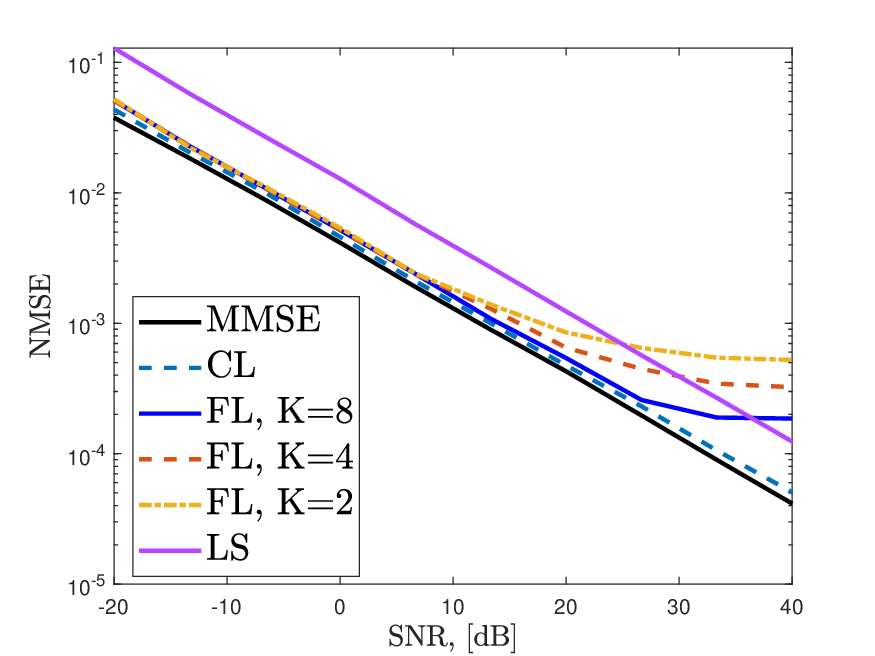}\label{fig_U_NMSE_all} } 
			\caption{ Validation RMSE (a) and channel estimation NMSE (b)  with respect to $K$ in massive MIMO scenario.	}
			\label{fig_U_Test}
		\end{figure}
		
		\subsection{Computational Complexity}
		
		We further examine the computational complexity of the proposed CNN architecture. The time complexity of the convolutional layers can be written as~\cite{deepCNN_ChannelEstimation,elbir2020withoutCSI}
		\begin{align}
		\mathcal{C}_\mathrm{CL} = 
		\mathcal{O}\bigg( \sum_{l=1}^{N_\mathrm{CL}}D_x^{(l)}D_y^{(l)} W_x^{(l)}W_y^{(l)}N_\mathrm{SF}^{(l-1)}N_\mathrm{SF}^{(l)} \bigg),
		\end{align}
		where $D_x^{(l)}, D_y^{(l)}$ are the column and row sizes of each output feature map, $ W_x^{(l)},W_y^{(l)}$ are the 2-D filter size of the $l$-th layer. $N_\mathrm{SF}^{(l-1)}$ and $N_\mathrm{SF}^{(l)}$ denote the number of input and output feature maps of the $l$-th layer respectively. Table~\ref{tableConvLayerParameters} lists the parameters of each convolutional layer for an $N_\mathrm{MS}\times N_\mathrm{BS}\times3$ input. Thus, the complexity of three convolutional layers with $128$@$3\times 3$ spatial filters approximately becomes
		\begin{align}
		\mathcal{C}_\mathrm{CL} \approx \mathcal{O}\big( 3\cdot 9 \cdot 128^2N_\mathrm{MS}N_\mathrm{BS} \big).
		\end{align}
		The time complexity of the fully connected layer similarly is
		\begin{align}
		\mathcal{C}_\mathrm{FCL}= \mathcal{O}\bigg( D_x D_y \kappa N_\mathrm{FCL}  \bigg), 
		\end{align}
		where $N_\mathrm{FCL}= 1024$ is the number of units with $\kappa = 1/2$ dropout. $ D_x=128N_\mathrm{MS}N_\mathrm{BS}$ and $D_y=1$ are the 2-D input size of the  fully connected layer respectively. Then, the time complexity of the fully connected layer approximately is 
		\begin{align}
		\mathcal{C}_\mathrm{FCL} \approx  \mathcal{O} \big(4\cdot 128^2 N_\mathrm{MS}N_\mathrm{BS} \big).
		\end{align}
		Hence the total time complexity of \textsf{ChannelNet} is $\mathcal{C} = \mathcal{C}_\mathrm{CL} + \mathcal{C}_\mathrm{FCL}$, which approximately is
		\begin{align}
		\mathcal{C} \approx \mathcal{O}\big( 3\cdot 9 \cdot 128^2N_\mathrm{MS}N_\mathrm{BS} +  4\cdot 128^2 N_\mathrm{MS}N_\mathrm{BS}) \big),
		\end{align}
		which is further simplified as $ \approx \mathcal{O}\big(31\cdot 128^2 N_\mathrm{MS}N_\mathrm{BS} \big)$. {\color{black}Since the computation of the pseudo-inverse of the received pilot data is required in the testing stage, the complexity order of LS and MMSE estimation are $\mathcal{O}\big(N_\mathrm{MS}^2N_\mathrm{BS}^2 \big)$ and $\mathcal{O}\big(N_\mathrm{MS}^3N_\mathrm{BS}^3 \big)$, respectively~\cite{deepCNN_ChannelEstimation,elbir2020TL}}.
		
		\textcolor{black}{Fig.~\ref{fig_Complexity} shows the time complexity comparison of CNN, MMSE and LS with respect to $N_\mathrm{MS}N_\mathrm{BS}$. We see that \textsf{ChannelNet} has higher complexity than LS. As the number of antennas, i.e., $N_\mathrm{MS}N_\mathrm{BS}$ increases, the complexity of MMSE becomes closer to that of \textsf{ChannelNet}, it becomes larger after approximately $N_\mathrm{MS}N_\mathrm{BS}\geq 720$. While the complexity of \textsf{ChannelNet} seems comparable with the conventional techniques, it is able to run more efficiently by using parallel processor, e.g., GPUs, which can significantly reduce the computation time~\cite{deepCNN_ChannelEstimation,elbir2020TL,elbir2019online}. However,  the implementation with GPUs is not straightforward for the other algorithms, and it requires algorithm-dependent processor configuration. } 
		
	}

	\begin{figure}[t]
		\centering
		\subfloat[]{\includegraphics[draft=false,width=.5\columnwidth]{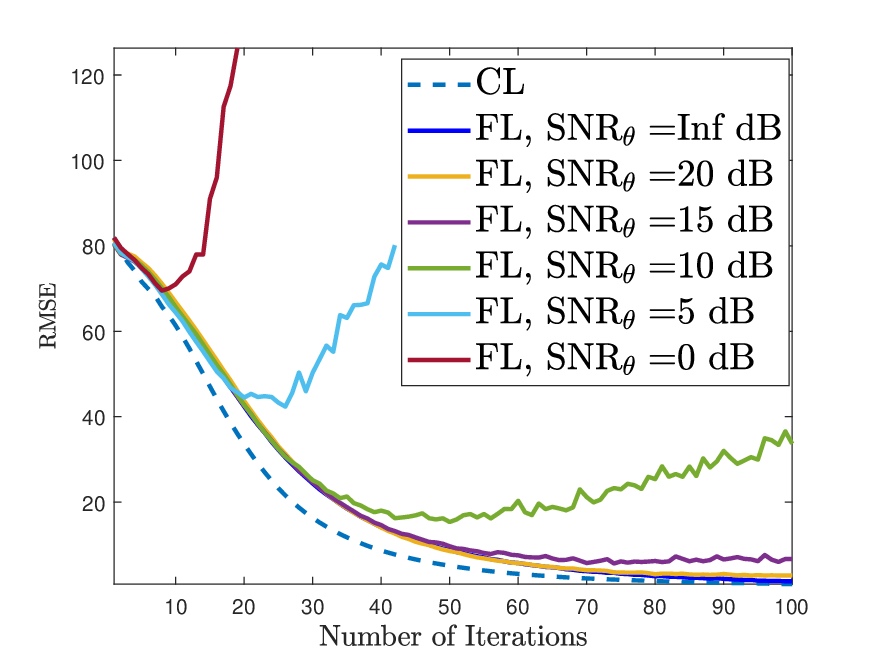}\label{fig_SNR_all} } 
		\subfloat[]{\includegraphics[draft=false,width=.5\columnwidth]{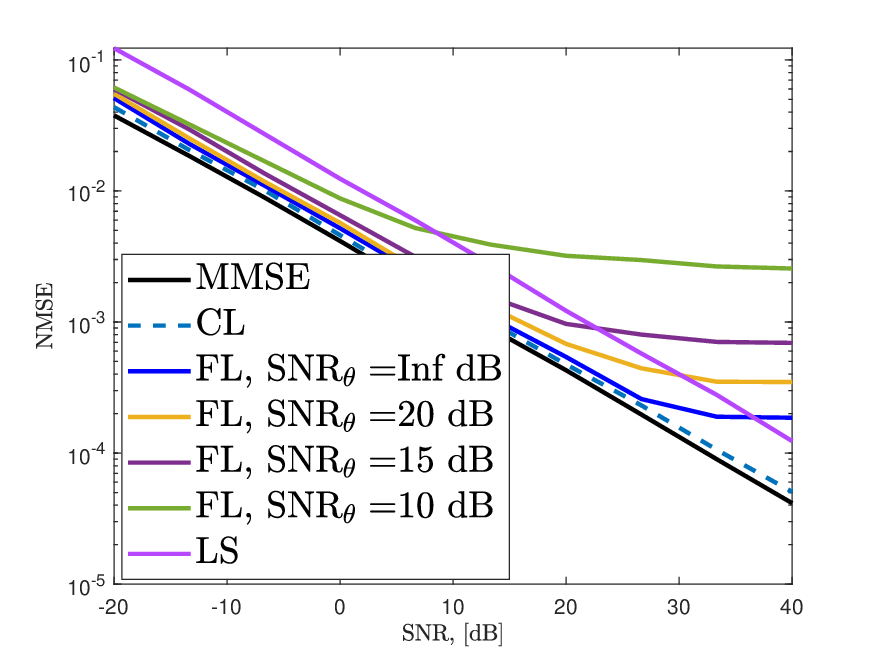} \label{fig_SNR_NMSE_all}} 
		\caption{ Validation RMSE (a) and channel estimation NMSE (b)  with respect to  $\mathrm{SNR}_{\boldsymbol{\theta}}$ in massive MIMO scenario, respectively.	}
		\label{fig_SNR_Test}
	\end{figure}

	\begin{figure}[t]
		\centering
		\subfloat[]{\includegraphics[draft=false,width=.5\columnwidth]{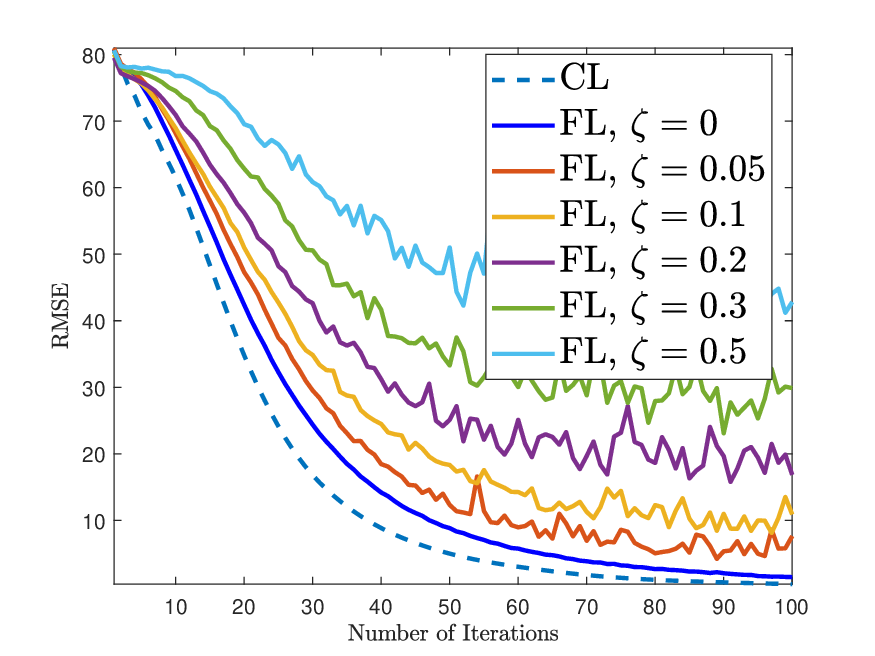}\label{fig_SNRimpulse_all} } 
		\subfloat[]{\includegraphics[draft=false,width=.5\columnwidth]{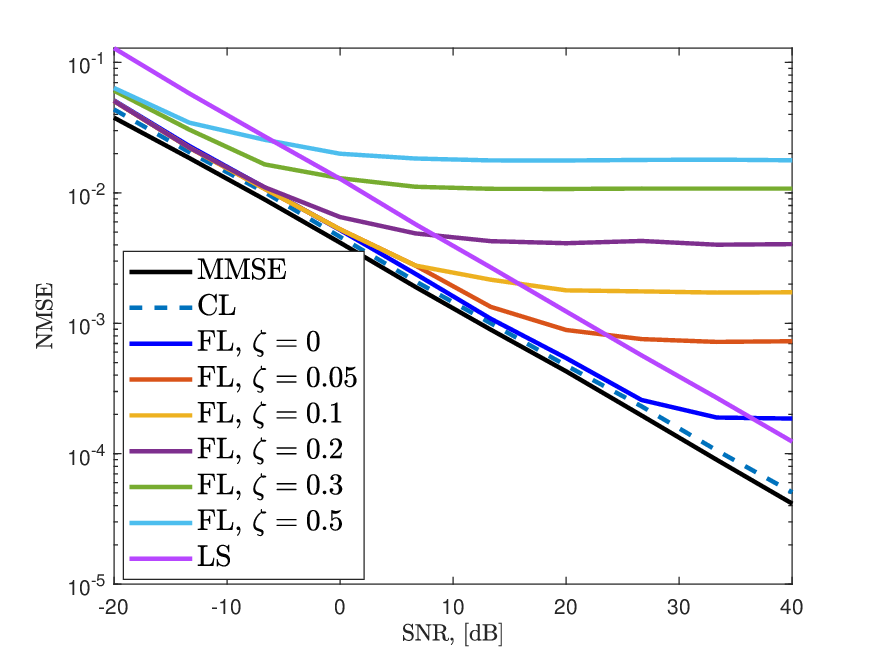} \label{fig_SNRimpulse_NMSE_all}} 
		\caption{ Validation RMSE (a) and channel estimation NMSE (b)  with respect to  $\zeta \in [0, 0.5]$ in massive MIMO scenario.	}
		\label{fig_SNRimpulse_Test}
	\end{figure}

	\section{Numerical Simulations}
	\label{sec:Sim}
	The goal of the simulations is to compare the performance of the proposed FL-based channel estimation approach to the channel estimation performance of the state-of-the-art ML-based channel estimation techniques SF-CNN~\cite{deepCNN_ChannelEstimation} and MLP~\cite{deepLearningChannelAndDOAEst}, and the MMSE and LS estimation in terms of normalized MSE (NMSE), defined by $\mathrm{NMSE} = \frac{1}{J_T KM} \sum_{i=1}^{J_T} \sum_{k=1}^{K}\sum_{m=1}^{M}$ $ \frac{\| \mathbf{H}_k[m] - \hat{\mathbf{H}}_k^{(i)}[m] \|_\mathcal{F}^2}{\| \mathbf{H}_k[m]  \|_\mathcal{F}^2   } ,$	where $J_T = 100$ number of Monte Carlo trials. We also present the validation RMSE of the training process, defined by $	 \mathrm{RMSE} =\left(\frac{1}{|\mathcal{D}_\mathrm{val}|}  \sum_{i = 1}^{|\mathcal{D}_\mathrm{val}|}\| f( \widetilde{\mathcal{X}}^{(i)}|\boldsymbol{\theta}) - \widetilde{\mathcal{Y}}^{(i)}  \|_\mathcal{F}^2\right)^{1/2},$	where $\widetilde{\mathcal{X}}^{(i)}$ and $\widetilde{\mathcal{Y}}^{(i)}$ respectively denote the input-output pairs in the validation dataset $\mathcal{D}_\mathrm{val}$, which includes $20\%$ of the whole dataset $\mathcal{D}$, hence, we have $|\mathcal{D}_\mathrm{val}| = 0.2 |\mathcal{D}|$.
	
	The local dataset of each user includes $N=100$ different channel realizations for $K=8$ users. The number of antennas in the massive MIMO scenario at the BS and users are $N_\mathrm{BS}=128$ and $N_\mathrm{MS}=32$, respectively, and we select $M=16$ and $L=5$. For the RIS-assisted scenario, $N_\mathrm{BS} = N_\mathrm{RIS} = 64$. Hence, we have the same number of input elements for both scenario, i.e, $128\cdot 32 = 64 \cdot 64$. In both scenarios, location of each user is selected as $\phi_{k,l}\in \Phi_k$ and  $ \bar{\phi}_{k,l} \in \bar{\Psi}_k$, where $\Phi_k$ and $\bar{\Psi}_k$ are the equally-divided subregions of the angular domain $\Theta$, i.e.,  $\Theta =\bigcup_{k\in \mathcal{K}} \Phi_k = \bigcup_{k\in \mathcal{K}}\bar{\Psi}_k$, respectively. The pilot data are generated as  $\overline{\mathbf{S}} = \mathbf{I}_{M_\mathrm{BS}}$ and $\overline{\mathbf{S}}_\mathrm{RIS} = \mathbf{I}_{M_\mathrm{BS}}$ for $M_\mathrm{BS} = N_\mathrm{BS}$ and $M_\mathrm{MS} = N_\mathrm{MS}$. We selected $\overline{\mathbf{F}}[m]$ and $\overline{\mathbf{W}}[m]$ as the first $M_\mathrm{BS}$ columns of an $N_\mathrm{BS}\times N_\mathrm{BS}$ DFT matrix and the first $M_\mathrm{MS}$ columns of an $N_\mathrm{MS}\times N_\mathrm{MS}$ DFT matrix, respectively~\cite{deepCNN_ChannelEstimation}. During training, we have added AWGN on the input data for three SNR levels, i.e., SNR$=\{20, 25, 30\}$ dB,  for $G_\mathrm{mMIMO}=20$ and $G_\mathrm{RIS}= 20M$ realizations in order to provide robust performance against noisy input~\cite{elbirDL_COMML,elbir_LIS} in both scenarios. As a result, both training datasets have the same number of input-output pairs as  $\textsf{D}_\mathrm{mMIMO} = 3MKNG_\mathrm{mMIMO} = 3\cdot16\cdot8\cdot100\cdot20=768,000$ and $\textsf{D}_\mathrm{RIS} = 3KNG_\mathrm{RIS} = 3\cdot8\cdot100\cdot320=768,000$, respectively. The proposed \textsf{ChannelNet} model is realized and trained in MATLAB on a PC with a $2304$-core GPU. For CL, we use the SGD algorithm with momentum of $0.9$ and the mini-batch size $M_B = 128$,  and  update the network parameters with learning rate $0.001$. For FL, we train \textsf{ChannelNet} for $T=100$ iterations/rounds. Once the training is completed, the labels of the validation data (i.e., $20\%$ of the whole dataset) are used in prediction stage. During the prediction stage, each user estimates its own channel by feeding \textsf{ChannelNet} with $\mathbf{G}_k[m]$ ($\boldsymbol{\Upsilon}_k$) and obtains $\hat{\mathbf{H}}_k[m]$ ($\hat{\mathbf{h}}_{\mathrm{B},k}$ and $\hat{\mathbf{V}}_k$) at the output for massive MIMO (RIS) scenario, respectively\footnote{The source codes of the FL-based channel estimation scheme can be found at \texttt{https://sites.google.com/view/elbir/publications}.}.

	\begin{figure}[t]
		\centering
		\subfloat[]{\includegraphics[draft=false,width=.5\columnwidth]{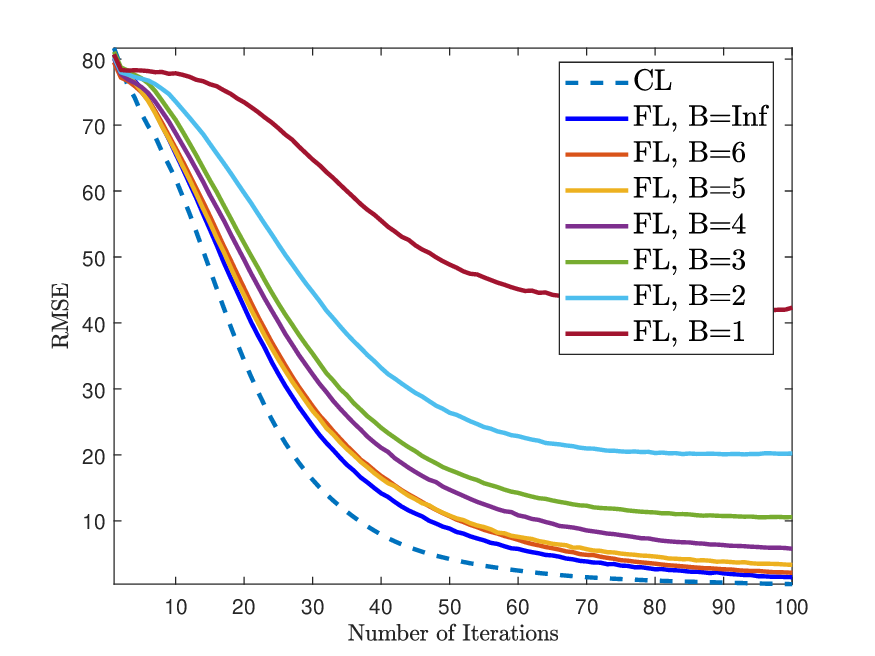} } 
		\subfloat[]{\includegraphics[draft=false,width=.5\columnwidth]{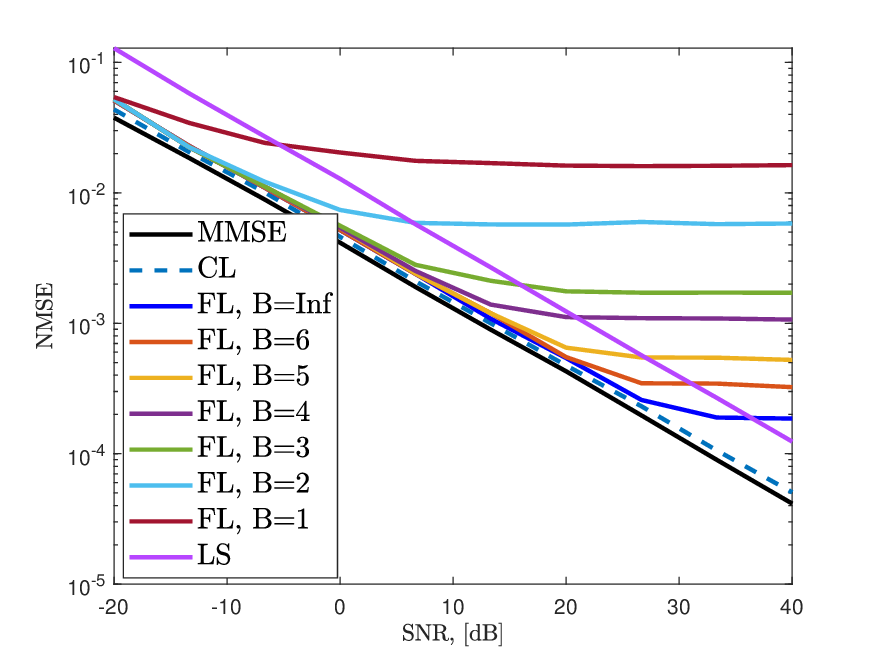} } 
		\caption{Validation RMSE (a) and   channel estimation NMSE (b) for different quantization levels in massive MIMO scenario, respectively.
		}
		\label{fig_QuantizationTest}
	\end{figure}
	\begin{figure}[h]
		\centering
		{\includegraphics[draft=false,width=.9\columnwidth]{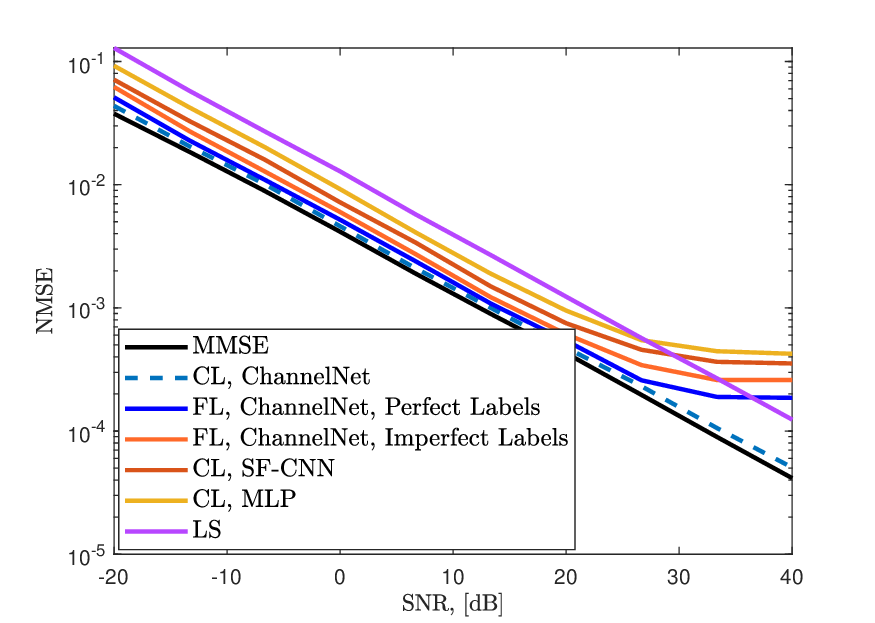} } 
		\caption{Channel estimation NMSE for different algorithms in  massive MIMO scenario.
		}
		\label{fig_NMSE_ALL}
	\end{figure}

	\begin{figure}[h]
		\centering
		{\includegraphics[draft=false,width=.9\columnwidth]{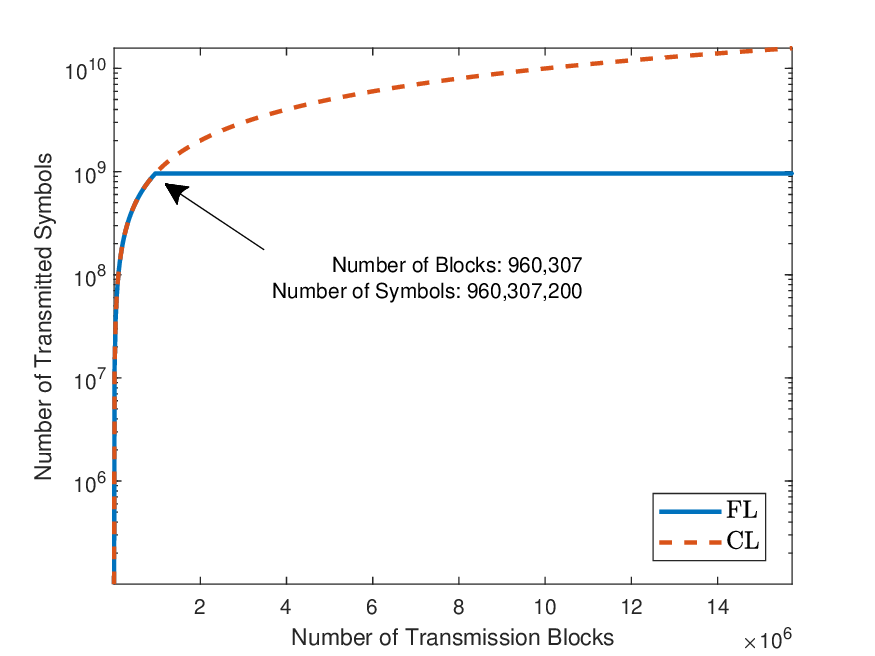} } 
		\caption{Communication overhead for FL- and CL-based model training.
		}
		\label{fig_transmissionOverhead}
	\end{figure}

	\subsection{Channel Estimation in Massive MIMO}

	\begin{figure}[t]
		\centering
		\subfloat[]{\includegraphics[draft=false,width=.5\columnwidth]{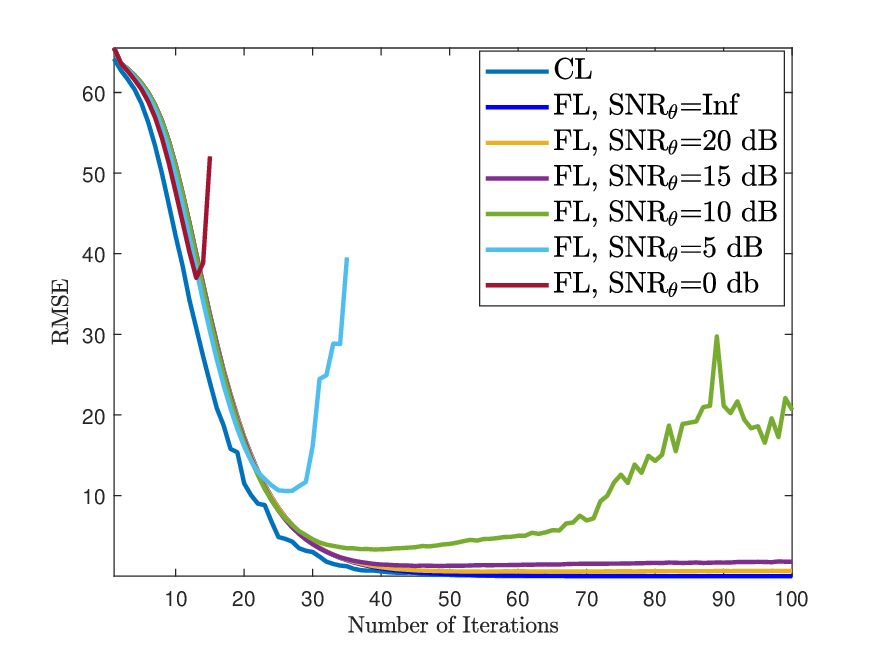}\label{fig_SNR_all_RIS} } 
		\subfloat[]{\includegraphics[draft=false,width=.5\columnwidth]{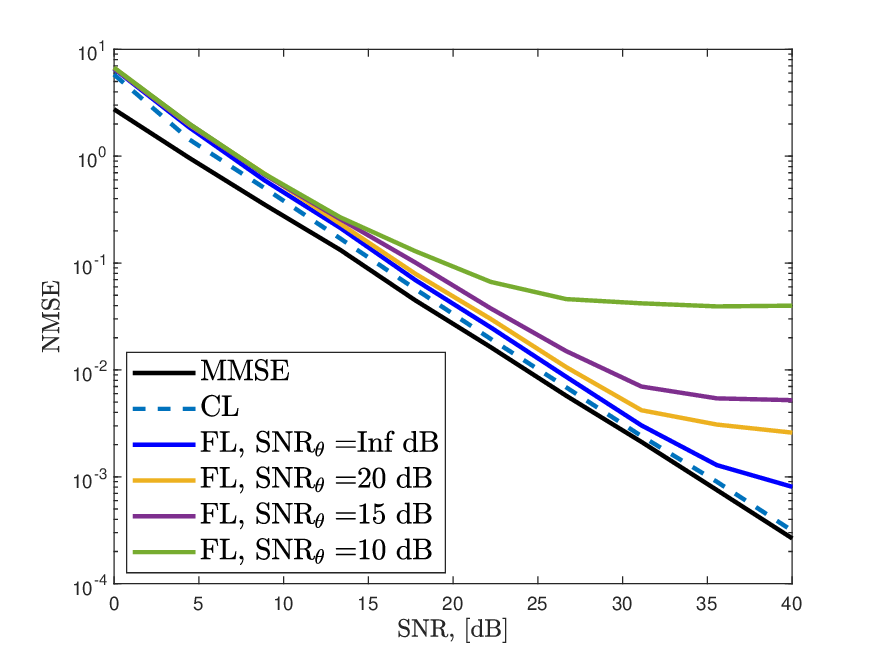} \label{fig_SNR_NMSE_all_RIS}} 
		\caption{ Validation RMSE (a) and channel estimation NMSE (b)  with respect to  $\mathrm{SNR}_{\boldsymbol{\theta}}$ in RIS-assisted massive MIMO scenario, respectively.	}
		\label{fig_SNR_Test_RIS}
	\end{figure}

	In Fig.~\ref{fig_U_Test}, we present the training performance (Fig.~\ref{fig_U_all}) and the channel estimation NMSE (Fig.~\ref{fig_U_NMSE_all}) of the proposed FL approach for channel estimation for different number  of users. In this scenario, we fix the total dataset size $\textsf{D}$ by selecting $G = 20\cdot \frac{8}{K}$. As $K$ increases, the training performance is observed to improve and gets closer to the performance of CL since the model updates superposed at the BS become more robust against the noise. As $K$ decreases, the corruptions in the model aggregation increase due to the diversity in the training dataset. 
	
	Fig.~\ref{fig_SNR_Test} shows the training and channel estimation performance for different noise levels added to the transmitted gradient and model data when $K=8$. Here, we add AWGN onto both $\mathbf{g}_k(\boldsymbol{\theta}_t)$ and $\boldsymbol{\theta}_t$ with respect to $\mathrm{SNR}_{\boldsymbol{\theta}}$, where  $\mathrm{SNR}_{\boldsymbol{\theta}} = 20\log_{10}\frac{||\mathbf{g}_k(\boldsymbol{\theta}_t)||_2^2}{\sigma_{\boldsymbol{\theta}}^2  }  $. We observe in Fig.~\ref{fig_SNR_all} that the training diverges for low $\mathrm{SNR}_{\boldsymbol{\theta}}$ (e.g., SNR$_{\boldsymbol{\theta}}\leq 5$ dB) due to the corruptions in the model parameters. The corresponding channel estimation performance is presented in Fig.~\ref{fig_SNR_NMSE_all} when the \textsf{ChannelNet} converges  and at least SNR$_{\boldsymbol{\theta}}\leq 15$ dB is required to obtain reasonable channel estimation performance, e.g., $\mathrm{NMSE}\leq 0.001$. 
	
	Fig.~\ref{fig_SNRimpulse_Test} shows the training and channel estimation performance in case of an impulsive noise causing the loss of gradient and model data. In this experiment, we multiply $\mathbf{g}_k(\boldsymbol{\theta}_t)$ and $\boldsymbol{\theta}_t$  with $\mathbf{u}\in \mathbb{R}^P$ as $\mathbf{g}_k(\boldsymbol{\theta}_t)\odot\mathbf{u}$ and $\boldsymbol{\theta}_t\odot\mathbf{u}$, where  the $\lfloor 100\zeta \rfloor$ elements of $\mathbf{u}$ are $0$ and the remaining terms are $1$. This allows us to simulate the case when a portion of the gradient/model data are completely lost during transmission. We observe that the loss of model data significantly affects both training and channel estimation accuracy. Therefore, reliable channel estimation demands at most $5\%$ parameter loss during transmission.

	\begin{figure}[t]
		\centering
		\subfloat[]{\includegraphics[draft=false,width=.5\columnwidth]{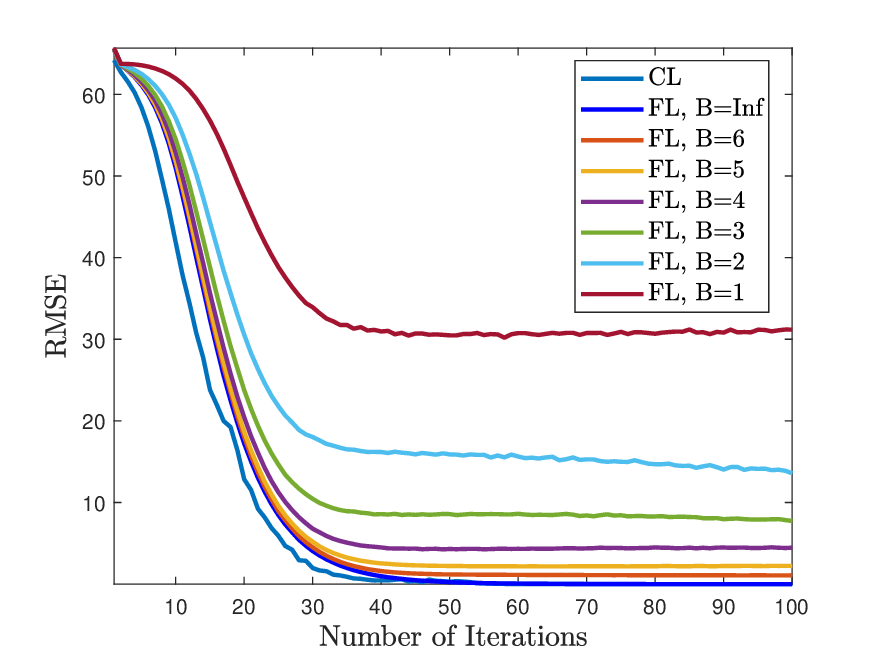} } 
		\subfloat[]{\includegraphics[draft=false,width=.5\columnwidth]{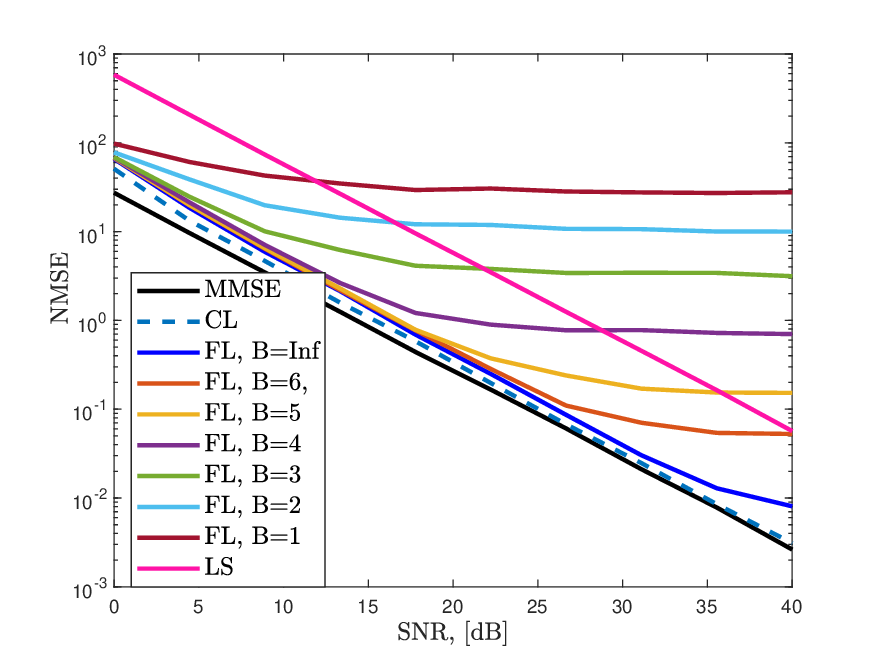} } 
		\caption{Validation RMSE (a) and channel estimation NMSE (b) for different quantization levels in RIS-assisted massive MIMO, respectively.
		}
		\label{fig_QuantizationTestRIS}
	\end{figure}

	Fig.~\ref{fig_QuantizationTest} shows the training and channel estimation performance when the transmitted data (i.e., $\mathbf{g}_k(\boldsymbol{\theta}_t)$ and $\boldsymbol{\theta}_t$) are quantized with $B$ bits. As expected, the performance improves as $B$ increases and  at least $B=5$ bits are required to obtain a reasonable channel estimation performance. Compared to the results in Fig.~\ref{fig_SNR_Test}, quantization has more influence on the accuracy than  $\mathrm{SNR}_{\boldsymbol{\theta}}$.

	In Fig.~\ref{fig_NMSE_ALL}, we present the channel estimation NMSE for different algorithms when $K=8$. We train \textsf{ChannelNet} with both CL and FL frameworks and observe that CL follows the MMSE performance closely. CL provides better performance than that of FL since it has access the whole dataset at once. Nevertheless, FL has  satisfactory channel estimation performance despite  decentralized training. Specifically, FL and CL have similar NMSE for SNR$\leq 25$ dB and the  performance of FL maxes out in high SNR regime. This is because the learning model loses  precision due to FL training and cannot perform better. This is a common problem in ML-based techniques~\cite{deepCNN_ChannelEstimation,elbir_LIS}. In order to improve the performance, over-training can be employed so that more precision can be obtained. However, this introduces overfitting, i.e., the model memorizes the data, hence, it cannot perform well for different inputs. In Fig.~\ref{fig_NMSE_ALL}, the comparison between the training with perfect (true channel data) and imperfect (estimated channel via ADCE) labels is also presented. The use of imperfect labels causes a slight performance degradation, while still providing less NMSE than SF-CNN and MLP.  The other algorithms also exhibit similar behavior but  perform worse than \textsf{ChannelNet}. This is because SF-CNN and MLP have  convolutional-only and fully-connected-only layers, respectively. In contrast, \textsf{ChannelNet} includes both structures, hence, exhibiting better feature extraction and data mapping performance.

	According to the analysis in Sec.~\ref{sec:DataComp}, {\color{black}the communication overhead of FL and CL are $2PTK = 2\cdot 600,192 \cdot 100 \cdot 8 \approx 960\times 10^6$ and $ (3N_\mathrm{MS} N_\mathrm{BS} + 2N_\mathrm{MS}N_\mathrm{BS}) \textsf{D} = (5\cdot 128\cdot 32)768,000 \approx 16\times 10^{9}$, respectively. This clearly shows the effectiveness of FL over CL.} We also present the number of transmitted symbols during training with respect to data transmission blocks in Fig.~\ref{fig_transmissionOverhead}, where we assume that $1000$ data symbols are transmitted at each transmission block. We can see that, it takes about $1\times 10^6$ data blocks to complete the gradient/model transmission in FL (see, e.g., Fig.~\ref{fig_DiagramFL}) whereas CL-based training demands approximately $16\times 10^{6}$ data blocks to complete the task for training dataset transmission (see, e.g., Fig.~\ref{fig_DiagramML}). Therefore, the communication overhead of FL is approximately $16$ times lower than that of CL.

	\subsection{Channel Estimation in RIS-assisted Massive MIMO}
	In Fig.~\ref{fig_SNR_Test_RIS}, we present the validation RMSE and the channel estimation NMSE. We compute the NMSE of both direct channel and the cascaded channel together and present the results in a single plot. Similar results are obtained for model training, which diverges when $\mathrm{SNR}_{\boldsymbol{\theta}} \leq 5$ dB and channel estimation NMSE becomes relatively small if $\mathrm{SNR}_{\boldsymbol{\theta}}\geq 15 $ dB.

	Fig.~\ref{fig_QuantizationTestRIS} shows the validation RMSE and channel estimation NMSE for different quantization levels. The small number of bits causes the loss of precision in channel estimation NMSE. Similar to the massive MIMO scenario, at least $B\geq 5$ bits are required to obtain satisfactory channel estimate performance at large SNRs, i.e., $\mathrm{SNR} \geq20 $dB.



	\section{Conclusions}
	\label{sec:Conc}
	In this paper, we propose a FL framework for channel estimation in conventional and RIS-assisted massive MIMO systems. We evaluate the performance of the proposed approach via several numerical simulations for different number of users and when the gradient/model parameters are quantized and corrupted by noise. We show that at least $5$ bit quantization and $15$ dB SNR on the model parameters are required for reliable channel estimation performance, i.e, $\mathrm{NMSE}\leq 0.001$. We further analyze the scenario when a portion of the gradient/model parameters are completely lost and observe that FL exhibits satisfactory performance under at most $5\%$ information loss. We also examine the channel estimation performance of the proposed CNN architecture with both perfect and imperfect labels. A slight performance degradation is observed in case of imperfect labels as compared to the perfect CSI case. Nevertheless, the performance of imperfect label scenario strongly depends on the accuracy of the channel estimation algorithm employed during training dataset collection. Furthermore, the proposed CNN architecture provides lower NMSE than the state-of-the-art NN architectures.  Apart from the channel estimation performance, FL-based approach enjoys approximately $16$ times lower transmission overhead as compared to the CL-based training. \textcolor{black}{As a future work, we plan to develop compression-based techniques for both training data and the model parameters to further reduce the communication overhead.}
	{\color{black}
		\appendices
		\section{Proof of Theorem 1}
		
		\label{appendix1}
		We first make the following assumptions needed to ensure the convergence, which are typical for the $\l_2$-norm regularized linear regression, logistic regression, and softmax classifiers~\cite{fl_convergenceOnNIIDData,robustFL,fl_By_Google}. 
		
		\textit{Assumption 1:} The loss function $\mathcal{L}(\boldsymbol{\theta})$ is convex, i.e., $\mathcal{L}((1 - \lambda)\boldsymbol{\theta} + \lambda \boldsymbol{\theta}') \leq (1 - \lambda)\boldsymbol{\theta} \mathcal{L}(\boldsymbol{\theta}) + \lambda\mathcal{L}(\boldsymbol{\theta}')$ for $\lambda \in [0,1]$ and arbitrary  $\boldsymbol{\theta}$ and $\boldsymbol{\theta}'$.
		
		\textit{Assumption 2:} $\mathcal{L}(\boldsymbol{\theta})$ is \textit{L-Lipschitz}, i.e., $||\mathcal{L}(\boldsymbol{\theta}) - \mathcal{L}(\boldsymbol{\theta}')|| \leq  L ||\boldsymbol{\theta} - \boldsymbol{\theta}' ||$ for arbitrary  $\boldsymbol{\theta}$ and $\boldsymbol{\theta}'$.
		
		\textit{Assumption 3:} $\mathcal{L}(\boldsymbol{\theta})$ is \textit{$\beta$-Smooth}, i.e., $||\nabla\mathcal{L}(\boldsymbol{\theta}) - \nabla\mathcal{L}(\boldsymbol{\theta}')|| \leq  \beta ||\boldsymbol{\theta} - \boldsymbol{\theta}' ||$ for arbitrary  $\boldsymbol{\theta}$ and $\boldsymbol{\theta}'$.
		
		In order to prove Theorem 1, we first investigate the $\beta$-\textit{Smoothness} of $\bar{\mathcal{L}}(\boldsymbol{\theta})$ in  the following lemma.
		
		\textit{Lemma 1:} $\bar{\mathcal{L}}(\boldsymbol{\theta})$ is a $\bar{\beta}$-\textit{Smooth} function with $||\nabla \bar{\mathcal{L}}(\boldsymbol{\theta}) - \nabla\bar{\mathcal{L}}(\boldsymbol{\theta}')|| \leq  \bar{\beta} ||\boldsymbol{\theta} - \boldsymbol{\theta}' ||$, where $\bar{\beta} = (1 + \sigma_{\Delta}^2) \beta$.
		
		\textit{Proof:} Using (\ref{lossFLRegularized}), we get
		\begin{align}
		&||\nabla \bar{\mathcal{L}}(\boldsymbol{\theta}) - \nabla\bar{\mathcal{L}}(\boldsymbol{\theta}')||  \nonumber \\
		&= || \nabla (\mathcal{L}(\boldsymbol{\theta})+ \sigma_{\Delta}^2||\nabla \mathcal{L}(\boldsymbol{\theta}) ||^2 ) \nonumber \\
		& \hspace{50pt} - \nabla (\mathcal{L}(\boldsymbol{\theta}')+ \sigma_{\Delta}^2||\nabla \mathcal{L}(\boldsymbol{\theta}')||^2)  ||\nonumber \\
		&=||  \big(\nabla \mathcal{L}(\boldsymbol{\theta}) + \sigma_{\Delta}^2\nabla||\nabla \mathcal{L}(\boldsymbol{\theta}) ||^2 \big) \nonumber \\
		&\hspace{50pt}- \big(\nabla \mathcal{L}(\boldsymbol{\theta}') + \sigma_{\Delta}^2\nabla||\nabla \mathcal{L}(\boldsymbol{\theta}') ||^2\big)  || \nonumber \\
		& =|| \nabla {\mathcal{L}}(\boldsymbol{\theta}) - \nabla{\mathcal{L}}(\boldsymbol{\theta}')  + \sigma_{\Delta}^2 \nonumber \\
		&\hspace{0pt} \times \big(\nabla \mathrm{tr}\{\nabla{\mathcal{L}}(\boldsymbol{\theta})^\textsf{T} \nabla {\mathcal{L}}(\boldsymbol{\theta})   \}  - \nabla \mathrm{tr}\{\nabla{\mathcal{L}}(\boldsymbol{\theta}')^\textsf{T} \nabla {\mathcal{L}}(\boldsymbol{\theta}')   \}\big)   || \nonumber \\
		& = ||  \nabla {\mathcal{L}}(\boldsymbol{\theta}) -\nabla {\mathcal{L}}(\boldsymbol{\theta}')   + \sigma_{\Delta}^2 \big(   \nabla {\mathcal{L}}(\boldsymbol{\theta}) - \nabla{\mathcal{L}}(\boldsymbol{\theta}')  \big)     || \nonumber \\
		& = || (1 + \sigma_{\Delta}^2) \big(\nabla {\mathcal{L}}(\boldsymbol{\theta}) -\nabla {\mathcal{L}}(\boldsymbol{\theta}')\big) || \nonumber \\
		&=(1 + \sigma_{\Delta}^2) || \nabla {\mathcal{L}}(\boldsymbol{\theta}) -\nabla {\mathcal{L}}(\boldsymbol{\theta}') ||.\label{eq6}
		\end{align}
		By incorporating (\ref{eq6}), Assumption 2 and  $1 + \sigma_{\Delta}^2 \geq 0$, we get
		\begin{align}
		\label{lemmaResult}
		||\nabla \bar{\mathcal{L}}(\boldsymbol{\theta}) - \nabla\bar{\mathcal{L}}(\boldsymbol{\theta}')|| \leq  \bar{\beta} ||\boldsymbol{\theta} - \boldsymbol{\theta}' ||^2,
		\end{align}
		where $\bar{\beta} = (1 + \sigma_{\Delta}^2) \beta$. \qed


		Using (\ref{lemmaResult}),  Assumption 2 and Assumption 3 imply that $\bar{\mathcal{L}}(\boldsymbol{\theta})$ is second order differentiable as $\nabla^2 \bar{\mathcal{L}}(\boldsymbol{\theta})\preceq \bar{\beta} \mathbf{I}_P $. Using this fact, performing a quadratic expression around $\bar{\mathcal{L}}(\boldsymbol{\theta})$ yields  
		\begin{align}
		\label{eq:quadraticExp}
		\bar{\mathcal{L}}(\boldsymbol{\theta}') &\leq \bar{\mathcal{L}}(\boldsymbol{\theta}) + \nabla \bar{\mathcal{L}}(\boldsymbol{\theta})^\textsf{T} (\boldsymbol{\theta}' - \boldsymbol{\theta})  + \frac{1}{2} \nabla^2 \bar{\mathcal{L}}(\boldsymbol{\theta}) ||\boldsymbol{\theta}' - \boldsymbol{\theta} ||^2 \nonumber \\
		&\leq \mathcal{L}(\boldsymbol{\theta}) + \nabla \bar{\mathcal{L}}(\boldsymbol{\theta})^\textsf{T} (\boldsymbol{\theta}' - \boldsymbol{\theta})  + \frac{1}{2} \bar{\beta}  ||\boldsymbol{\theta}' - \boldsymbol{\theta} ||^2.
		\end{align}
		Substituting the GD update $ \boldsymbol{\theta}' = \boldsymbol{\theta} - \eta \nabla \bar{\mathcal{L}}(\boldsymbol{\theta}) $ in (\ref{eq:quadraticExp}), we get
		\begin{align}
		\bar{\mathcal{L}}(\boldsymbol{\theta}') &\leq \bar{\mathcal{L}}(\boldsymbol{\theta}) + \nabla \bar{\mathcal{L}}(\boldsymbol{\theta})^\textsf{T} (\boldsymbol{\theta}' - \boldsymbol{\theta})  + \frac{1}{2} \bar{\beta}  ||\boldsymbol{\theta}' - \boldsymbol{\theta} ||^2 \nonumber \\
		& = \bar{\mathcal{L}}(\boldsymbol{\theta})  + \nabla \bar{\mathcal{L}}(\boldsymbol{\theta})^\textsf{T} (\boldsymbol{\theta} - \eta \nabla \bar{\mathcal{L}}(\boldsymbol{\theta})  - \boldsymbol{\theta}) \nonumber \\
		& \hspace{10pt}+ \frac{1}{2} \nabla^2 \bar{\mathcal{L}}(\boldsymbol{\theta}) ||\boldsymbol{\theta} - \eta \nabla \bar{\mathcal{L}}(\boldsymbol{\theta}) - \boldsymbol{\theta} ||^2 \nonumber \\
		& = \bar{\mathcal{L}}(\boldsymbol{\theta})  - \eta \nabla \bar{\mathcal{L}}(\boldsymbol{\theta})^\textsf{T} \nabla\bar{\mathcal{L}}(\boldsymbol{\theta}) +  \frac{1}{2} \bar{\beta}  ||\eta \nabla \bar{\mathcal{L}}(\boldsymbol{\theta}) ||^2  \nonumber \\
		&= \bar{\mathcal{L}}(\boldsymbol{\theta})  - \eta || \nabla \bar{\mathcal{L}}(\boldsymbol{\theta})||^2 + \frac{1}{2} \bar{\beta} \eta^2 || \nabla \bar{\mathcal{L}}(\boldsymbol{\theta})||^2 \nonumber \\
		& = \bar{\mathcal{L}}(\boldsymbol{\theta}) - (1 - \frac{\bar{\beta} \eta }{2}) \eta  || \nabla \bar{\mathcal{L}}(\boldsymbol{\theta})||^2, \label{useConvexity}
		\end{align}
		which bounds the GD update $\bar{\mathcal{L}}(\boldsymbol{\theta}')$ with $\bar{\mathcal{L}}(\boldsymbol{\theta})$. Now, let us bound $\bar{\mathcal{L}}(\boldsymbol{\theta}')$ with the optimal objective value $\bar{\mathcal{L}}(\boldsymbol{\theta}_\star)$. Using Assumption 1, we have
		\begin{align}
		\label{dueToConvexity}
		\bar{\mathcal{L}}(\boldsymbol{\theta}_\star )& \geq \bar{\mathcal{L}}(\boldsymbol{\theta})  + \nabla \bar{\mathcal{L}}(\boldsymbol{\theta})^\textsf{T} (\boldsymbol{\theta}_\star - \boldsymbol{\theta}), \nonumber \\
		\bar{\mathcal{L}}(\boldsymbol{\theta} )& \leq  \bar{\mathcal{L}}(\boldsymbol{\theta}_\star)  + \nabla \bar{\mathcal{L}}(\boldsymbol{\theta})^\textsf{T} (\boldsymbol{\theta} - \boldsymbol{\theta}_\star).
		\end{align}
		Furthermore, using $\eta \leq \frac{1}{\bar{\beta}}$, we have $-(1 - \frac{\bar{\beta}\eta}{2}) = \frac{1}{2}\bar{\beta}\eta - 1 \leq \frac{1}{2 }\bar{\beta} (1/\bar{\beta}) -1 = \frac{1}{2} - 1 = -\frac{1}{2}$. Thus, (\ref{useConvexity}) becomes
		\begin{align}
		\label{useConvexity2}
		\bar{\mathcal{L}}(\boldsymbol{\theta}') \leq \bar{\mathcal{L}}(\boldsymbol{\theta}) - \frac{\eta}{2}  || \nabla \bar{\mathcal{L}}(\boldsymbol{\theta})||^2
		\end{align}
		By plugging (\ref{dueToConvexity}) into (\ref{useConvexity2}), we get 
		\begin{align}
		&\bar{\mathcal{L}}(\boldsymbol{\theta}') \leq \bar{\mathcal{L}}(\boldsymbol{\theta}_\star) + \nabla \bar{\mathcal{L}}(\boldsymbol{\theta})^\textsf{T} (\boldsymbol{\theta} 
		- \boldsymbol{\theta}_\star ) - \frac{\eta}{2}  || \nabla \bar{\mathcal{L}}(\boldsymbol{\theta})||^2,
		\end{align}
		which can be rewritten as
		\begin{align}
		\label{eq1}
		\bar{\mathcal{L}}(\boldsymbol{\theta}') \hspace{-3pt}-\hspace{-3pt} \bar{\mathcal{L}}(\boldsymbol{\theta}_\star) \hspace{-3pt}\leq \hspace{-3pt}
		\frac{1}{2\eta} \big( 2\eta \nabla \bar{\mathcal{L}}(\boldsymbol{\theta})^\textsf{T} (\boldsymbol{\theta} \hspace{-3pt}
		- \boldsymbol{\theta}_\star )\hspace{-3pt} - \hspace{-3pt}\eta^2 ||\hspace{-3pt} \nabla \bar{\mathcal{L}}(\boldsymbol{\theta})||^2  \big).
		\end{align}
		By adding $\frac{1}{2\eta}(||\boldsymbol{\theta} - \boldsymbol{\theta}_\star   ||^2 - ||\boldsymbol{\theta} - \boldsymbol{\theta}_\star   ||^2)$ into the right hand side of (\ref{eq1}), we get
		\begin{align}
		\label{eq2}
		\bar{\mathcal{L}}(\boldsymbol{\theta}')\hspace{-3pt} -\hspace{-3pt} \bar{\mathcal{L}}(\boldsymbol{\theta}_\star) \hspace{-3pt}\leq \hspace{-3pt}
		\frac{1}{2\eta} \big( ||\boldsymbol{\theta}\hspace{-3pt} - \boldsymbol{\theta}_\star   ||^2\hspace{-3pt}  - \hspace{-3pt}||\boldsymbol{\theta} - \boldsymbol{\theta}_\star\hspace{-3pt} -\hspace{-3pt} \eta \nabla \bar{\mathcal{L}}(\boldsymbol{\theta}) ||^2  \big),
		\end{align}
		which is obtained after incorporating the expansion of $||\boldsymbol{\theta} - \boldsymbol{\theta}_\star - \eta \nabla \bar{\mathcal{L}}(\boldsymbol{\theta}) ||^2$. Substituting the GD update $\boldsymbol{\theta}' = \boldsymbol{\theta} - \eta\nabla\bar{\mathcal{L}}(\boldsymbol{\theta}) $ into (\ref{eq2}), we have 
		\begin{align}
		\bar{\mathcal{L}}(\boldsymbol{\theta}') - \bar{\mathcal{L}}(\boldsymbol{\theta}_\star) \leq \frac{1}{2\eta } \bigg(||\boldsymbol{\theta} - \boldsymbol{\theta}_\star   ||^2 - ||\boldsymbol{\theta}' - \boldsymbol{\theta}_\star   ||^2   \bigg).
		\end{align}
		Now, replacing $\boldsymbol{\theta}'$ by $\boldsymbol{\theta}_i$ and summing over $i = 1,\dots, t$ yield
		\begin{align}
		&\sum_{i = 1}^{t} (\bar{\mathcal{L}}(\boldsymbol{\theta}_i) - \bar{\mathcal{L}}(\boldsymbol{\theta}_\star))  \hspace{-3pt} \leq \hspace{-3pt}\sum_{i = 1}^{t} \frac{1}{2\eta } \bigg(||\boldsymbol{\theta}_{i-1}\hspace{-3pt} - \boldsymbol{\theta}_\star   ||^2 \hspace{-3pt}- ||\boldsymbol{\theta}_i - \boldsymbol{\theta}_\star   ||^2   \bigg) \nonumber \\
		& = \frac{1}{2\eta }\bigg(||\boldsymbol{\theta}_0 - \boldsymbol{\theta}_\star   ||^2 - ||\boldsymbol{\theta}_t - \boldsymbol{\theta}_\star   ||^2   \bigg) \hspace{-3pt}\leq  \frac{1}{2\eta }||\boldsymbol{\theta}_0 - \boldsymbol{\theta}_\star  ||^2,   \label{eq4}
		\end{align}
		where the summation on the right hand side disappears since the consecutive terms cancel each other. Since $\bar{\mathcal{L}}(\boldsymbol{\theta}_t)$ is a decreasing function, we have 
		\begin{align}
		\label{eq3}
		\bar{\mathcal{L}}(\boldsymbol{\theta}_t) - \bar{\mathcal{L}}(\boldsymbol{\theta}_\star) \leq  \frac{1}{t} \sum_{i=1}^t(\bar{\mathcal{L}}(\boldsymbol{\theta}_i) - \bar{\mathcal{L}}(\boldsymbol{\theta}_\star)).
		\end{align}
		Inserting (\ref{eq4}) into (\ref{eq3}), we finally have
		\begin{align}
		\bar{\mathcal{L}}(\boldsymbol{\theta}_t) - \bar{\mathcal{L}}(\boldsymbol{\theta}_\star) \leq \frac{1}{2\eta t}||\boldsymbol{\theta}_0 - \boldsymbol{\theta}_\star  ||^2 .
		\end{align}
		\qed
		
	}
	
	\bibliographystyle{IEEEtran}
	\footnotesize{\bibliography{IEEEabrv,references_070_journal}}

	\begin{IEEEbiographynophoto} {Ahmet M. Elbir} (IEEE Senior Member) received the Ph.D. degree from Middle East Technical University in 2016 He is a Senior Researcher at Duzce University, Duzce, Turkey, and Research Fellow at the University of Hertfordshire, Hatfield, UK.
	\end{IEEEbiographynophoto}
	\begin{IEEEbiographynophoto} {Sinem Coleri} (IEEE Senior Member) received the Ph.D. degree from
		the University of California at Berkeley in	2005. She is	a Faculty Member with the Department of Electrical and Electronics Engineering, Koc University, Turkey. 
	\end{IEEEbiographynophoto}

\end{document}